\documentclass[traditabstract]{aa}  
\usepackage{graphicx}
\usepackage{color}
\usepackage{txfonts}
\usepackage{epsfig}
\usepackage{natbib}
\usepackage{subfigure}
\usepackage{enumerate}
\usepackage{longtable}
\usepackage{morefloats}

\newcommand{\cii}{{\ion{C}{II}}}

\newcommand{\mgii}{{\ion{Mg}{II}}}
\newcommand{\siiv}{{\ion{Si}{IV}}}
\newcommand{\oiv}{{\ion{O}{IV}}}
\newcommand{\siv}{{\ion{S}{IV}}}
\newcommand{\feii}{{\ion{Fe}{II}}}
\newcommand{\si}{{\ion{S}{I}}}
\newcommand{\fex}{{\ion{Fe}{X}}}

\newcommand{\ovii}{{\ion{O}{VII}}}
\newcommand{\ov}{{\ion{O}{V}}}

\begin{document}

\title{Density diagnostics derived from the \ion{O}{IV} and \ion{S}{IV}
 intercombination lines observed by IRIS}

\author{V. Polito \inst{1}
\and
G. Del Zanna\inst{1}
\and 
J. Dud\'ik\inst{2}
\and
H. E. Mason\inst{1}
\and
A. Giunta\inst{3}
\and
K.K. Reeves\inst{4}
}

\institute{Department of Applied Mathematics and Theoretical Physics, CMS, University of Cambridge, Wilberforce Road,\\
	      Cambridge CB3 0WA, United Kingdom
              \email{vp323@cam.ac.uk}
         \and Astronomical Institute, Academy of Sciences of the Czech Republic, 25165 Ond\v{r}ejov, Czech Republic,
         \and STFC Rutherford Appleton Laboratory, Chilton, Didcot, Oxon. OX11 0QX, UK,
         \and Harvard-Smithsonian Center for Astrophysics, 60 Garden Street, Cambridge MA 01238, USA}


  \abstract{The intensity of the \oiv~2s$^{2}$ 2p $^{2}$P-2s2p$^{2}$ $^{4}$P and \siv~3 s$^{2}$ 3p $^{2}$P- 3s 3p$^{2}$ $^{4}$ P intercombination lines around 1400~\AA~observed with the \textit{Interface Region Imaging Spectrograph} (IRIS) provide a useful tool to diagnose the electron number density ($N_\textrm{e}$) in the solar transition region plasma. We measure the electron number density in a variety of solar features observed by IRIS, including an active region (AR) loop, plage and brightening, and the ribbon of the 22-June-2015 M 6.5 class flare. By using the emissivity ratios of  \oiv\ and \siv\ lines, we find that our observations are consistent with the emitting plasma being near isothermal (log$T$[K] $\approx$ 5) and iso-density ($N_\textrm{e}$ $\approx$~10$^{10.6}$ cm$^{-3}$) in the AR loop. Moreover, high electron number densities ($N_\textrm{e}$ $\approx$~10$^{13}$ cm$^{-3}$) are obtained during the impulsive phase of the flare by using the \siv\ line ratio. We note that the \siv\ lines provide a higher range of density sensitivity than the \oiv\ lines. Finally, we investigate the effects of high densities ($N_\textrm{e}$ $\gtrsim$ 10$^{11}$ cm$^{-3}$) on the ionization balance. In particular, the fractional ion abundances are found to be shifted towards lower temperatures for high densities compared to the low density case. We also explored the effects of a non-Maxwellian electron distribution on our diagnostic method. }
  
\keywords{
Sun: transition region, Sun: UV radiation, techniques: spectroscopic, atomic data}

\maketitle
\begin{figure*}[!htbp]
	\centering
	\includegraphics[width=0.8\textwidth]{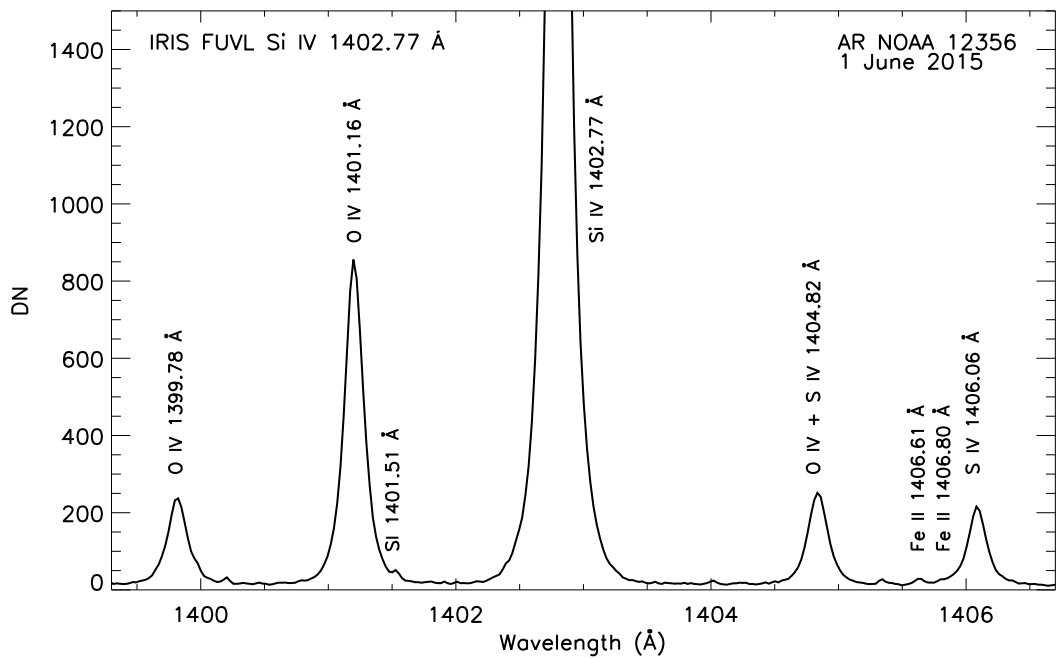} 	
      \caption{Example of a spectrum observed by IRIS in the AR 12356 (see Sect. \ref{Sect:4.2}) showing emission lines from the transition region. The figure shows the \oiv\ + \siv\ blend at around 1404.82~\AA, and the wings of the \oiv\ 1401.16~\AA~and \siv\ 1406.06~\AA~lines blending with photospheric cooler transitions (\si\ and \feii\ respectively). In addition, we note that the shape of the IRIS line profiles for these ions is not Gaussian, but rather similar to a generalized Lorentzian function. }
      \label{Fig:IRIS_sp}
  \end{figure*}

\section{Introduction}
\label{Sect:1}
The intercombination lines from \oiv\ and \siv\ around 1400~\AA~provide useful electron density ($N_\mathrm{e}$) diagnostics in a variety of solar features and astrophysical plasmas \citep[see, e.g.,][]{Flower75,Feldman79,Bhatia80}. These transitions are particularly suitable for performing density measurements  as their ratios are known to be largely independent of the electron temperature, and only weakly dependent on the electron distribution \citep{Dudik14}. The other advantage is that these lines are close in wavelength, minimizing any instrumental calibration effects. 


However, discrepancies between theoretical ratios and observed values have been reported in the past by several authors. For instance, \citet{Cook95} calculated emission line ratios from different \oiv\ and \siv\ line pairs by using solar observations from the \textit{High Resolution Telescope Spectrograph} (HRTS) and the SO82B spectrograph on board \textit{Skylab} as well as stellar observations from the \textit{Hubble Space Telescope}. They found that the observed ratios from \oiv\ and \siv\ would imply electron densities which differed significantly with each other (by up to an order of magnitude). Some of the discrepancies were subsequently identified by \citet{Keenan02} as due to line blends and low accuracy in the atomic data calculations. They obtained more consistent density diagnostics from \oiv\ and \siv\ ratios by using updated atomic calculations together with observations from SOHO/SUMER. 
 Nevertheless, some inconsistencies still remained \citep{DelZanna02}. 

There is  now renewed interest in the literature concerning these transitions, because some of the \oiv\ and \siv\ intercombination lines, together with the  \siiv\ resonance lines, are routinely observed with the \textit{Interface Region Imaging Spectrograph} \citep[IRIS;][]{DePontieu14} at much higher spectral, spatial and temporal resolution than previously. For example, \cite{peter_etal:2014} used the intensities of the \oiv\ vs. \siiv\ lines to propose that very high densities, on the order of 10$^{13}$ cm$^{-3}$ or higher, are present in the so-called IRIS plasma `bombs'. 
Line ratios involving an \oiv\ forbidden transition and a \siiv\ allowed transition have been used in the past to provide electron densities during solar flares and transient brightenings \citep[e.g.,][]{Cheng81,Hanssen81}. However, the validity of using \oiv\ to \siiv\ ratios has been hotly debated because these ratios gave very high densities compared to the more reliable ones obtained from the \oiv\ ratios alone \citep[see, e.g.,][]{hayes_shine:1987}. In addition, \cite{judge:2015} recalled several issues that should be taken into account when considering the \siiv /\oiv\ density diagnostic. The main ones were: 1) \oiv\ and \siiv\ ions are formed at quite different temperatures in equilibrium and hence a change in the \oiv\ / \siiv\ ratio could imply a change in the temperature rather than in the plasma density 2) the chemical abundances of O and Si are not known with any great accuracy and could be varying during the observed events 3) density effects on the ion populations could increase the \siiv\ / \oiv\ relative intensities by a factor of roughly  3--4. \cite{judge:2015} also mentioned the well-known problem of the "anomalous ions", e.g., the observed high intensities of the Li- and Na-like (as \siiv) ions \citep[see also][]{DelZanna02}.
Another important aspect to take into account is the effect of non-equilibrium conditions on the observed plasma diagnostics. It is well-known that strong variations in the line intensities are obtained when non-equilibrium ionisation is included in the numerical calculations \citep[see, e.g.,][]{Shen13,Raymond78,Mewe80,bradshaw_etal:04}. In particular, \cite{Doyle13} and \cite{Olluri13} investigated the consequences of time-dependent ionization on the formation of the \oiv\ and \siiv\ transition region lines observed by IRIS.  In addition, \cite{Dudik14} showed that non-Maxwellian electron distributions in the plasma can substantially affect the formation temperatures and intensity ratios of the IRIS \siiv\ and \oiv\ lines. These authors also suggested that the observing window used by IRIS should be extended to include \siv. Recent IRIS observation sequences have indeed included the \siv\ line near 1406~\AA. The \siv\ line ratios have a higher limit for density sensitivity than the \oiv\ line ratios and are thus particularly useful for diagnosing high densities which might occur in flares. Previous flare studies have in fact reported line ratios involving O ions which lay above the density sensitivity range, indicating an electron density in the excess of 10$^{12}$ cm$^{-3}$ \citep[e.g.,][]{Cook95,Polito16}. 

We present here the analysis of several IRIS observational datasets where \siiv, \oiv, and \siv\ lines were observed. We focus on the diagnostics based on the \oiv\ and \siv\ lines which we believe to be more reliable than those involving the \siiv\ to \oiv\ line ratios, because of the issues described above. 

The observations used in this work were obtained from a variety of solar features. Small spatial elements
were selected, to reduce multi-thermal and multi-density effects. Discrepancies in density diagnostics can in fact also arise if regions of plasma at different temperature and density are observed along the line of sight \citep{Doschek84,Almleaky89}. We discuss in some detail the various factors that affect the density measurements, and their uncertainties, as well as the different methods to obtain densities. 

We start by describing in Sect. \ref{Sect:2} the spectral lines and atomic data used in our analysis. We refer the reader to the Appendix \ref{Sect:A1} for a detailed review of the issues related to the atomic data and wavelengths for these lines.
 In Sect.~\ref{Sect:3} we briefly discuss the diagnostic methods. In Sect. ~\ref{Sect:4} we analyse a loop spectrum observed in Active Region (AR) NOAA 12356 on the 1 June 2015 and a spectrum acquired at the footpoints of the M 6.5 class flare on the 22 June 2015. The analysis of additional spectra observed in the AR can be found in Appendix \ref{Sect:A2}. Sect.\ref{Sect:5} presents a discussion on some of the physical processes which can affect the formation temperature of the ions studied in this work. Finally, the results of our analysis are discussed and summarized in Sect. \ref{Sect:6}.


\begin{figure}[!htbp]
	\centering
	\includegraphics[width=0.45\textwidth]{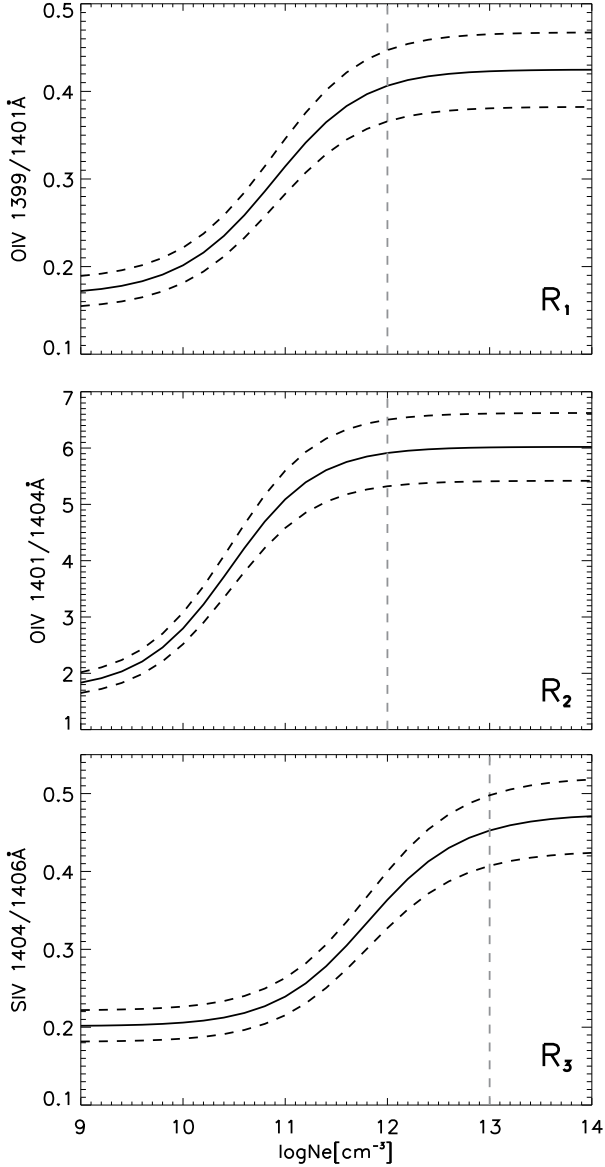} 	
      \caption{Theoretical ratios (continuous lines) of \oiv~1399/1401\AA, 1401/1404\AA~and 
\siv~1404/1406\AA~obtained by using the atomic data described in the text.
 The dotted curves show a $\pm$~10\% error for the theoretical ratios. 
The vertical lines indicate the high-density limit of  10$^{12}$ cm$^{-3}$ for the \oiv\ ratios and 10$^{13}$ cm$^{-3}$ for the \siv\ ratios.}
      \label{Fig:ratios}
  \end{figure}
\section{Transition-region lines observed by IRIS}
\label{Sect:2}
Tab.~\ref{tab:iris_lines} lists the \oiv, \siv\ and \siiv\ transition region (hereafter, TR) lines observed by the IRIS FUV spectrograph which have been analysed in this study. We note that in this context, TR indicates an approximate temperature of formation for emission lines rather than a layer of the solar atmosphere. An example of an IRIS AR spectrum showing these lines is given in Fig. \ref{Fig:IRIS_sp}. It is important to note that the IRIS spectral window has recently been extended to include the 3s$^2$ 3p $^2$P$_{3/2}$--3s 3p$^2$ $^4$P$_{5/2}$ \siv\ transition at 1406.06~\AA. 

The \oiv\ and \siv\ density-sensitive intensity ratios discussed in this work are: 

\begin{equation}
R_{1}=\frac{I_\oiv(1399.78~\AA)}{I_\oiv(1401.16~\AA)}
\label{eq:R1}
\end{equation}
\begin{equation}
R_{2}=\frac{I_\oiv(1401.16~\AA)}{I_\oiv(1404.81~\AA)}
\label{eq:R2}
\end{equation}
\begin{equation}
R_{3}=\frac{I_\siv(1404.85~\AA)}{I_\siv(1406.06~\AA)}
\label{eq:R3}
\end{equation}

The above ratios are shown in Fig.~\ref{Fig:ratios} as a function of electron density. They have been obtained by using the atomic data described in Sect. \ref{Sect:2.1}. The ratios $R_1$, $R_2$ and $R_3$ are useful density diagnostics in an intermediate density interval, between lower and higher densities (10$^{10}$--10$^{12}$ cm$^{-3}$ for \oiv\ and 10$^{11}$--10$^{13}$ cm$^{-3}$ for \siv). 

One limitation in the use of these lines is that they are normally observed to be extremely weak (the \oiv~ 1399.779~\AA\ line in particular). In addition, the variation of the $R_{\rm1}$ ratio over the density sensitivity range is not very large, and only spans around a factor of two in magnitude (top panel of  Fig.~\ref{Fig:ratios}), as already noted by e.g. \cite{feldman_doschek:1979}. The ratio $R_{\rm2}$ involving the $^{2}$P$_{3/2}$--$^{4}$P$_{3/2}$~\oiv\ component at 1404.81~\AA~is in principle better, but unfortunately this line is blended with the \siv\ 3$s^{2}$3p $^{2}$P$_{1/2}$--3s 3$p^{2}$ $^{4}$P$_{1/2}$ transition at 1404.85~\AA~(see Table~\ref{tab:iris_lines}). Various methods have been considered to address this issue in the past. \cite{Feldman79} discussed how the importance of the \siv\ to the \oiv\ line in the blend varies with the plasma density and temperature. For instance, at the quiet Sun density of 10$^{10}$ cm$^{-3}$, the \siv\ transition only contributes by around 10$\%$ to the 1404.82~\AA~(bl) line. The ratio between the \siv\ and \oiv\ blended lines is however significantly higher ($\approx$~50~$\%$) at densities of the order of 10$^{12}$ cm$^{-3}$. 

In this work, we estimate the intensity of the \oiv\ 1404.81~\AA~line by dividing the intensity of the unblended \oiv\ 1401.16~\AA~line with the value of the theoretical \textit{R$_{2}$} ratio at the density $N_\mathrm{e}$ which is determined by using the \textit{R$_{1}$} ratio:

\begin{equation}
I_{\oiv} (1404.81~\AA)=\frac{I_\oiv(1401.16~\AA)}{R_2}
\label{eq:Ioiv}
\end{equation}

The intensity of the blended \siv\ line at 1404.85~\AA~ is then simply obtained as the difference between the intensity of the 1404.82~\AA~blend and the estimated intensity of the \oiv\ 1404.81~\AA~line:
 
 \begin{equation}
I_{\siv} (1404.85~\AA)=I_{\oiv + \siv}(1404.82~\AA)-I_{\oiv} (1404.81~\AA)
\label{eq:Isiv}
\end{equation}

Finally, we compare the results obtained by using $R_{1}$ $R_{2}$, $R_{3}$ and the \siiv\ / \oiv\ ratio $R_{4}$:

\begin{equation}
R_\textrm{4}=\frac{I_\siiv(1402.77~\AA)}{I_\oiv(1401.16~\AA)}
\label{eq:R4}
\end{equation}

However, we bear in mind that this latter ratio is known to imply much higher densities than the $R_{1,2,3}$ ratios because of the issues outlined in the introduction \citep[see in particular][]{hayes_shine:1987}.
 
\begin{table}[!htbp]
\centering
\caption{\oiv\ and \siv\ transitions observed by IRIS. }
\begin{tabular}{llll}
 \hline\hline\noalign{\smallskip} 
$i-j$ &  Transition &  $\lambda$    & Ion   \\
& & \textbf{(\AA)}& \\
 \noalign{\smallskip}\hline\noalign{\smallskip} 
1--3 & 2s$^2$ 2p $^2$P$_{1/2}$ -- 2s 2p$^2$ $^4$P$_{1/2}$ &  1399.776 & \ion{O}{iv}  \\
2--5 & 2s$^2$ 2p $^2$P$_{3/2}$ -- 2s 2p$^2$ $^4$P$_{5/2}$ &  1401.163 & \ion{O}{iv}     \\
2--4 & 2s$^2$ 2p $^2$P$_{3/2}$ -- 2s 2p$^2$ $^4$P$_{3/2}$ &  1404.806 & \ion{O}{iv} (bl) \\
1--3 & 3s$^2$ 3p $^2$P$_{1/2}$ -- 3s 3p$^2$ $^4$P$_{1/2}$  &  1404.85 &  \ion{S}{iv} (bl)  \\ 
2--5 & 3s$^2$ 3p $^2$P$_{3/2}$ -- 3s 3p$^2$ $^4$P$_{5/2}$  &  1406.06  & \ion{S}{iv}   \\ 
1--2 & 3s $^2$S$_{1/2}$-- 3p $^2$P$_{1/2}$ & 1402.77 & \ion{Si}{iv} \\
\noalign{\smallskip}\hline
\label{tab:iris_lines}
\end{tabular}
\end{table}

\subsection{Atomic data, blends and wavelengths}
\label{Sect:2.1}

Various inconsistencies in the electron densities obtained from the \oiv~and \siv~ have been reported in the literature (see Appendix \ref{Sect:A1}). The main ones are removed with improved atomic data \citep[see][]{Keenan02}; however, some still remain, especially regarding the 1404.8~\AA\ blend \citep{DelZanna02}, which is about 30\% stronger than predicted. This discrepancy prompted a new \siv~calculation by the UK APAP network\footnote{www.apap-network.org}, presented in \cite{delzanna_badnell:2016}. Here, we use these \siv\ new atomic data, but we note that these data differ from the previous ones, available within the CHIANTI v.8 database \citep{delzanna_etal:2015_chianti_v8}, by only about 10\%. In particular, the excitation rates that drive the level population of the 3s 3p$^{2}$ $^{4}$P levels are within a few percent of those calculated by \cite{tayal:2000}, while variations in the $A$-values as calculated by different authors vary by at most 10\%. Therefore, we estimate an uncertainty in the intensity ratios of the \siv~intercombination lines to be about 10\%. This however means that the 30\% discrepancy in the 1404.8~\AA~blend is still present even if the new atomic data of \citet{delzanna_badnell:2016} are used.

For the \oiv\ transitions, we use the UK APAP network data \citep{liang_etal:2012} as distributed within the CHIANTI v.8 database \citep{delzanna_etal:2015_chianti_v8}. In Appendix \ref{Sect:A1}, we describe some of the earlier atomic data, and provide a Table of A-values, where we can see that variations are well within 10\%. Therefore, as in the \siv~ case, we estimate an overall uncertainty in the intensity ratios of the intercombination lines to be about 10\%. Finally, for the \siiv\ transition at 1402.77~\AA, we use the atomic data from CHIANTI v.8.

Another important issue is presented by possible blends in the emission lines under study. The \oiv~lines are sometimes observed to be blended with cooler emission lines, as described by \cite{Young15:report}. \citet{Polito16} noted that the \oiv~lines at 1399.78~\AA~and 1401.16~\AA~were blended with several photospheric lines, some of them unidentified, during the impulsive phase of an X class flare. \cite{Keenan02} suggested that \feii\ lines are blended with the \siv~1406.06~\AA~transition. The presence of two \feii\ lines (at 1405.61~\AA~and 1405.80~\AA) on the blue side of the \siv\ 1406~\AA~line is indeed confirmed in some of the observed spectra in this study, see Sect. \ref{Sect:4.1} and Fig. \ref{Fig:IRIS_sp}. We also note that in some previous observations \citep[e.g. SOHO/SUMER, see][]{Keenan02}, some of the lines were blended with second-order lines, an issue not present in the IRIS data. To avoid any source of error in our analysis, we carefully investigated the possibility of line blends in the observed \oiv\ and \siv\ spectra and included possible blends as an additional uncertainty in the line intensity measurement (see Sect. \ref{Sect:4.1}). 

Finally, we note that there has been some confusion in the literature regarding the wavelengths of the \oiv\ and \siv\ lines. For example, the wavelengths of the important \oiv\ and \siv\ blended lines around 1404.08~\AA\ have  been inverted in some cases. We list in Table~\ref{tab:iris_lines} our recommended wavelengths. The discussion on these wavelengths is quite involved and given in Appendix \ref{Sect:A1}. 

\begin{figure}[!ht]
	\centering
	\includegraphics[width=0.45\textwidth]{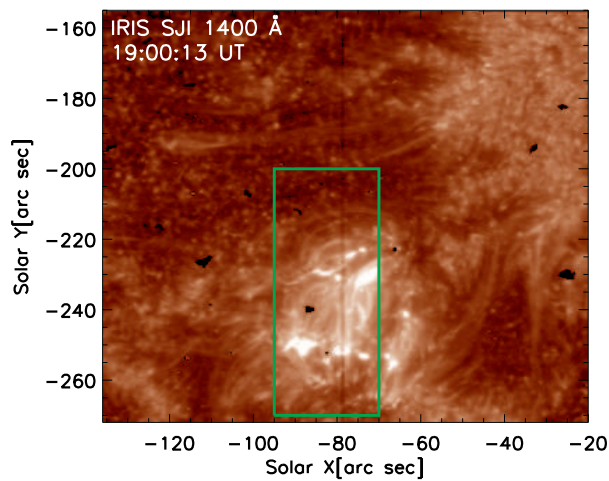} 	
      \caption{Overview of the AR NOAA 12356 as observed by the IRIS SJI in the 1400~\AA~passband at around 19:00~UT. See also the online movie 1 which shows the evolution of the SJI 1400~images over time. The green box overplot to the SJI image indicates the field of view of the IRIS spectrograph images in Fig. \ref{Fig:TR_maps}. }
      \label{Fig:SJI}
\end{figure}

\section{Density diagnostic method}
\label{Sect:3}
In this work, we estimate the electron density $N_\mathrm{e}$ of the emitting plasma by using an emissivity ratio method. This method is based on comparing the observed intensity $I_\mathrm{obs}$ and the calculated contribution function $G_\textrm{th}$($T,N_\textrm{e}$) of a set of optically thin spectral lines as a function of $N_\textrm{e}$ . 

The function $G_\textrm{th}$, in units of phot cm$^{3}$ s$^{-1}$, is defined as: 
\begin{equation}
	G_\textrm{th}(T,N_\textrm{e})=\frac{N(X^{+m}_j)}{N(X^{+m})}\frac{N(X^{+m})}{N(X)}A_\textrm{b}(X)\frac{N(H)}{N_\textrm{e}}\frac{A_{ji}}{N_\textrm{e}}
	\label{eq:Ith}
\end{equation}
where $N(X^{+m}_j) / N(X^{+m})$ is the population of the upper level $j$ relative to the total number density of the ion $X$$^{+m}$, calculated at a temperature $T$, and $N(X^{+m})/ N(X)$ is the relative abundance of the ion $X^{+m}$ in equilibrium. In addition, $A_{ji}$ is the spontaneous radiative transition probability, A$_\textrm{b}$(X) is the abundance of the element $X$ relative to hydrogen, and $N(H)/ N_\textrm{e}$ is taken as 0.83 in a completely ionized plasma. The product $A_{ji} N(X^{+m}_j) / N(X^{+m})$ is obtained by using the CHIANTI v8 \textit{IDL} routine \emph{emiss\_calc} for a given temperature $T$ and density $N_\textrm{e}$. We use photospheric chemical abundances $A_\textrm{b}(X)$ from \cite{asplund_etal:2009} and fractional ion abundances $N(X^{+m})/ N(X)$ from CHIANTI v8 and the Atomic Data and Analysis Structure (ADAS)\footnotemark[1]
\footnotetext[1]{www.adas.ac.uk,www.open.adas.ac.uk} databases.

Considering that the observed intensity $I_\textrm{obs}$ [phot s$^{-1}$ arcsec$^{-2}$ cm$^{-2}$] of a spectral line $\lambda_{ij}$ along the line of sight d$h$ can be expressed as:



%
\begin{equation}
	I_\textrm{obs}=\frac{hc}{4\pi\lambda_{ij}}\int_{h}G_\textrm{th}(T,N_\textrm{e})N_\textrm{e}^2 \mathrm{d}h \,,
	\label{eq:Iobs}
\end{equation}
the emissivity ratio $ER(T,N_\mathrm{e})$ of the spectral line can be defined as:
\begin{equation}
	ER(T,N_\textrm{e}) = \frac{I_\textrm{obs}}{G_\textrm{th}(T,N_\textrm{e})} \cdot C\,,
	\label{eq:ratio}
\end{equation}
where $C$ is a scaling constant depending on the geometry of the source and on the units used. It is the same for all lines originating in the same plasma source. The emissivity ratio curves presented in Sections \ref{Sect:4.2} and \ref{Sect:4.3} are normalized so that their y--axis ranges between 0 and 1.

For an almost iso-density plasma, the ratios $ER(T,N_\textrm{e})$ should then consistently intersect (within the errors) at the same value, giving the density $N_\textrm{e}$ of the emitting plasma source. This method can be applied to spectral lines from the \oiv\ and \siv\ ions, which are formed at close values of temperature and therefore, in a constant-pressure plasma, at similar densities. For instance, at formation temperatures of log$T$[K] $\approx$ 5.15 and 5.0 for \oiv\ and \siv\ respectively (as calculated in CHIANTI v.8 in ionization equilibrium), we expect the densities of two ions to be close within a factor of $\approx$ 1.8. This factor is close to the uncertainty, considering that even a small error in the observed ratio would produce a spread in the estimated density value using the theoretical ratio curves shown in Fig. \ref{Fig:ratios}.

The method described above is similar to the \textit{L}-function method suggested by \cite{Landi97}, with the main difference that they used a Differential Emission Measure (DEM) weighted temperature \textit{T$_0$} to calculate $G_\textrm{th}$. Given the lack of suitable spectral lines to perform a DEM analysis in our observations, we use an alternative approach to estimate the average plasma temperature $T$ (see Sect. \ref{Sect:4}). Except for a density factor, the emissivity ratio method used here is similar to the method presented by \citealt{DelZanna04}; however, these authors applied it to spectral lines formed within a single ion (\fex) only. Therefore, they could ignore the chemical and ion abundance factors in the definition of their emissivity ratio.

The advantage of using the emissivity ratio (Eq. \ref{eq:ratio}) compared to the individual intensity ratios $R_{1}$,$R_{2}$,$R_{3}$ is that each line $l$ is considered independently, thus allowing a better identification of the presence of possible discrepancies or anomalies in the intensity of a particular spectral line. 

It is important to note that in Eq. \ref{eq:ratio}, the $I_\textrm{obs}$ is also a function of $T$ and $N_\textrm{e}$ present in the emitting source (Eq. \ref{eq:Iobs}). This is especially the case if we are considering the blend at 1404.82~\AA. Here, the intensity of the individual \oiv~1404.81~\AA~and \siv~1404.85~\AA~lines will depend on the density and temperature assumed to deblend them (Eqs. \ref{eq:Ioiv}--\ref{eq:Isiv}). 

In order to test the validity of our method, we first simulated several \oiv~and \siv~synthetic spectra at known temperatures and densities using the atomic data described in Sect. \ref{Sect:2}. We then verified that by using the emissivity ratio method we were able to recover the same plasma temperature and density chosen for simulating the synthetic spectra. 

Finally, various issues need to be considered when comparing line intensities from different ions. For example, chemical abundances which are outlined below (Sect. \ref{Sect:3.1}). Some other issues will be discussed in Sect. \ref{Sect:5}.

\subsection{Chemical abundances}
\label{Sect:3.1}
It has long been observed that some solar TR and coronal features show variations in the 
chemical abundances, which are correlated with the 
first ionization potential (FIP) of the element \citep[see, e.g., ][]{Laming15}.
The low-FIP ($\leq$ 10 eV) elements are more abundant 
than the high-FIP ones, relative to the photospheric values (the FIP bias). 
Typical variations are of the order of 3--4. 
In our analysis, we have adopted the photospheric abundances recommended by 
\cite{asplund_etal:2009}. However, we note that with coronal abundances the \siiv\ 
intensity would increase compared to the \oiv\ intensity. S has an FIP of about 10,
but it normally shows intensity variations in line with those of the high-FIP elements
such as O, and therefore we do not expect large variations in the S/O abundance.
It is interesting to note that closed long-lived structures such as the 3~MK core loops in active regions show 
consistently an FIP bias of about 3 \citep{delzanna:2013_multithermal,delzanna_mason:2014}, 
while newly emerged regions normally show photospheric abundances. Unfortunately, IRIS does not observe enough TR lines to measure chemical abundances accurately.

 As we shall see in Sect. \ref{Sect:4}, there is a consistency in the relative 
intensities of the IRIS O and S lines, independent of the feature
observed. In particular, we have also examined the emission in the sites of chromospheric 
evaporation during the impulsive phase of flares, where we would normally
expect photospheric abundances. On the other hand, we find that the intensity of \siiv\ relative to \oiv\ and \siv\ is always very large and even increases in the flare spectra, which is the opposite of what we would expect if the variations were due to an elemental abundance variation. \siiv\ in fact behaves as an "anomalous ion" (see \citealt{DelZanna02} and references therein). Non-equilibrium effects such as non-equilibrium ionization \citep[e.g.,][]{Doyle13} and non-thermal electron distributions \citep[e.g.,][]{Dudik14} might be responsible for this discrepancy (see also Sect. \ref{Sect:5.2}.).


\begin{figure*}[!ht]
	\centering
	\includegraphics[width=\textwidth]{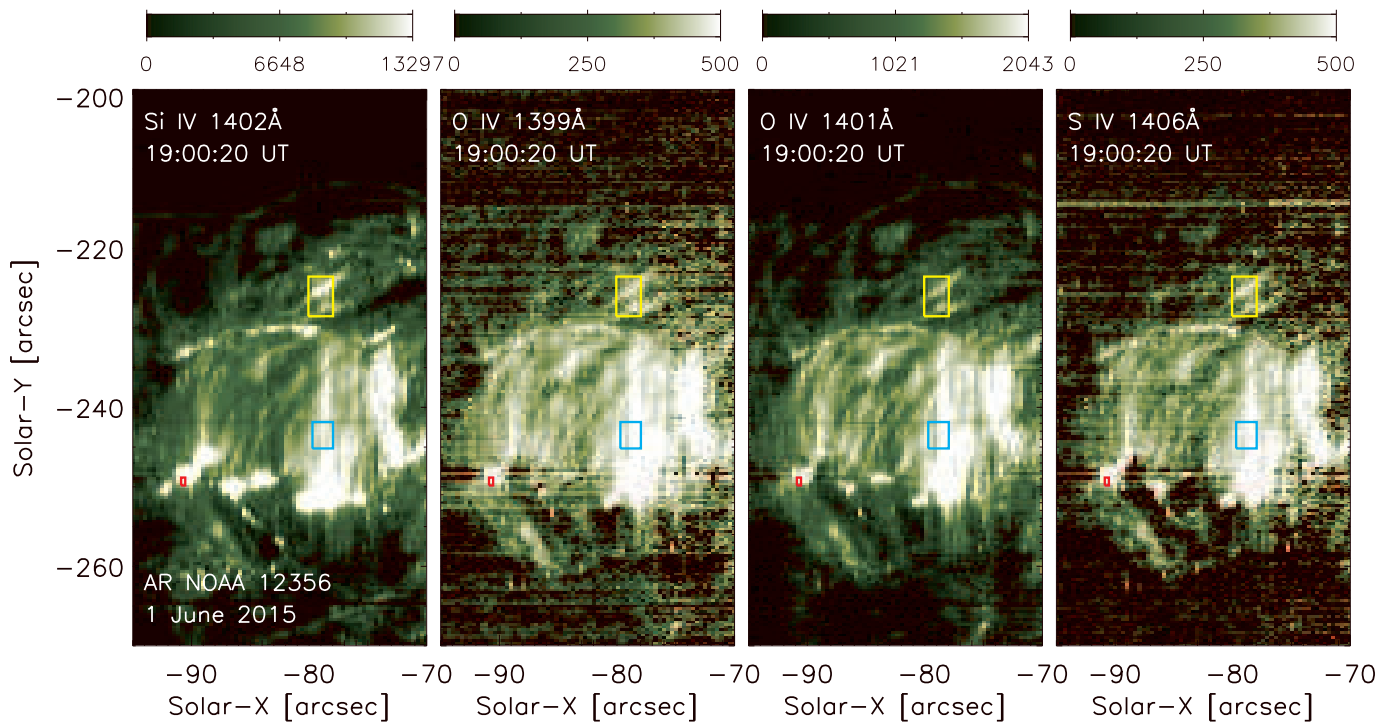} 	
      \caption{Monochromatic images of the AR NOAA 12356 observed by IRIS showing the intensity of different transition region spectral lines. From left to right: \siiv~1402.77~\AA~, \oiv~1399.78~\AA~and 1401.16~\AA~and \siv~1406.93~\AA. The units are phot s$^{-1}$ arcsec$^{-2}$ cm$^{-2}$. The yellow, light blue and red coloured boxes represent the regions where we acquired the plage, loop and bright point spectra respectively, which are shown in Fig. \ref{Fig:spectra}. The time halfway through the IRIS raster is indicated in each panel.}
      \label{Fig:TR_maps}
\end{figure*}

\begin{figure*}[!ht]
	\centering
	\includegraphics[width=\textwidth]{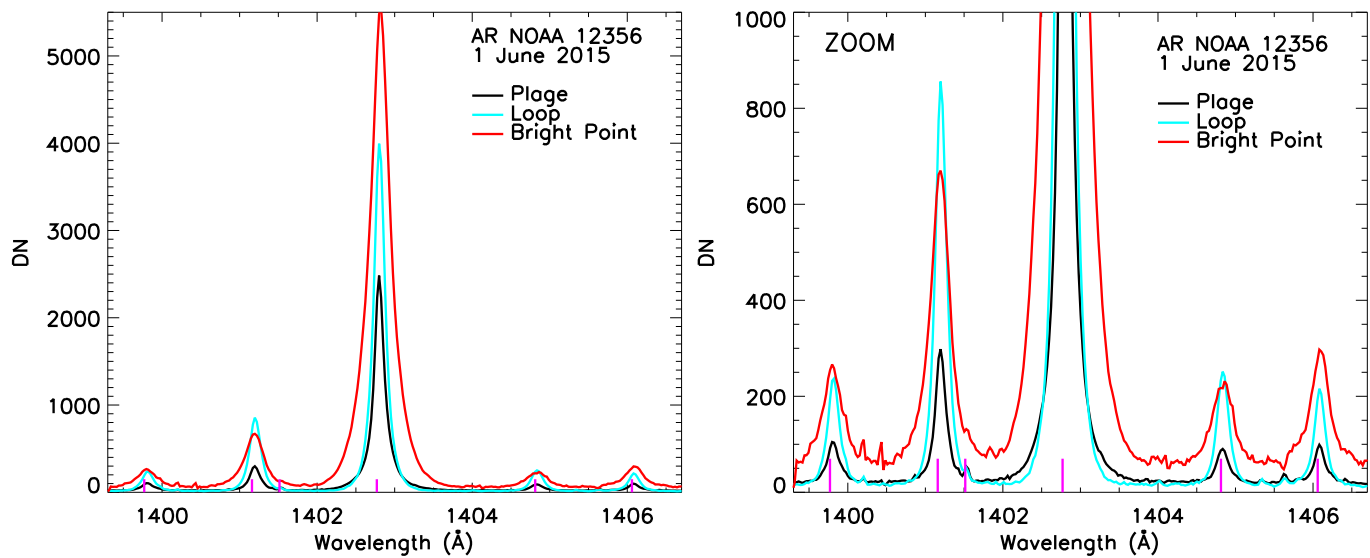} 	
      \caption{Spectra of the IRIS \siiv~1402.77~\AA~spectral window for a plage region (black), loop (light blue) and bright point (red). The spatial location where these spectra are acquired are shown by the coloured boxes in Fig. \ref{Fig:TR_maps} respectively. The pink vertical lines indicate the at-rest wavelength position for each of the spectral lines. The spectral lines do not show any significant Doppler shift apart from the $\approx$ 5-8 km s$^{-1}$ quiet-Sun systematic redshift.}
      \label{Fig:spectra}
\end{figure*}
\section{IRIS observations of an active region and a flare}
\label{Sect:4}
IRIS is a complex instrument with two major components. It acquires simultaneous spectra and images at very high temporal (up to 2s) and spatial (0.33$\arcsec$--0.4$\arcsec$) resolution. The IRIS spectrograph observes continua and emission lines over a very broad range of temperatures (log$T$[K]~=~3.7--7), including the TR lines under study (Tab. \ref{tab:iris_lines}). Simultaneously, the IRIS Slit Jaw Imager (SJI) provides high-resolution images in four different passbands (\cii~1330~\AA~, \siiv~1400~\AA, \mgii~k 2796~\AA~ and \mgii~wing 2830~\AA).

In this study, we examine the spectra of \oiv, \siv\ and \siiv\ lines observed by IRIS in the non-flaring AR NOAA 12356 on 1 June 2015 (see Fig. \ref{Fig:SJI}) and in AR 12371 during an M6.5 class flare on 22 June 2015.
We use level 2 data downloaded from the IRIS website \footnotemark[2], which are obtained from level 0 data after flat-field, geometry calibration and dark current subtraction, as detailed in the IRIS software notes \footnotemark[3]. In addition, we used the solarsoft routine $\emph{despik.pro}$ to remove the cosmic rays. The calibration of the wavelength scale was performed as described in \cite{Polito15,Polito16}.

 \footnotetext[2]{http://iris.lmsal.com/search/}
 \footnotetext[3]{http://iris.lmsal.com/documents.html}
 

We use IRIS spectroscopic data to measure the intensity $I_\textrm{obs}$ (expressed as data number (DN)) of the TR lines in different observed spectra. The intensities were then converted from DN to physical units (erg s$^{-1}$ sr$^{-1}$ cm$^{-2}$) by using the radiometric calibration detailed in the IRIS technical note 26~\footnotemark[3]. In particular, the updated version of the SolarSoft routine \emph{iris\_get\_response.pro} was used, which corrects the effective areas to take into account the instrumental degradation since launch. The errors associated to the intensities are derived as described in the following section. 
 
\subsection{Fitting the IRIS line profiles}
\label{Sect:4.1}
The line profiles observed in the AR under study (Sect. \ref{Sect:4.2}) present a non-Gaussian shape with narrow core and broad wings (Fig. \ref{Fig:spectra}). Such profiles could be fitted well with two Gaussian components (see Fig. \ref{Fig:spectra_appendix} in the Appendix \ref{Sect:A2}). We used the SolarSoft routine \textit{xcfit} to perform the fitting. The error estimated by the multi component fitting in the \textit{xcfit} routine takes into account the errors (as standard deviations) associated to the parameters of each Gaussian. Given the high number of free parameters in the double Gaussian fit, this error can be quite large, even though the sum of the two Gaussians approximates the observed spectrum well, as shown in Fig. \ref{Fig:spectra_appendix}. In order to have an alternative estimation of the errors associated to the observed intensities $I_\textrm{obs}$, we compare the calibrated intensities obtained by the fitting the line profiles with two Gaussians to the intensities obtained by summing the total intensity under the line profile. These errors are typically less than 10\% (see Tables \ref{tab:plage} and \ref{tab:bright} in Appendix \ref{Sect:A2}). In both method (fitting and sum) a background was subtracted by the spectral window before measuring the line intensities. This background can be estimated by averaging the counts over a spectral interval which is free of any emission lines.

 We note that the spectral lines in the AR plage and loop regions (see Fig. \ref{Fig:spectra} and Figs. \ref{Fig:spectra_appendix} in the Appendix \ref{Sect:A2}) present relatively narrow profiles which allows us to rule out the presence of significant blends, except the well-known blends of \si~1401.51~\AA~within the \oiv~1401.16~\AA~line and \oiv~1404.81~\AA~with \siv~1404.85~\AA~(see Sect. \ref{Sect:2}). In particular, in the plage spectrum (middle panel of Fig. \ref{Fig:spectra_appendix} in the Appendix \ref{Sect:A2}) we can clearly distinguish the two \feii\ lines (see Sect. \ref{Sect:2}) and the nearby the \siv\ 1406.06~\AA~line, ruling out any possible blending issue. For the plage and loop spectra, the difference between the summed and fitted intensities is very small (see Tables \ref{tab:loop}, \ref{tab:plage} and \ref{tab:bright} in the Appendix \ref{Sect:A2}); therefore we use the intensity values obtained by summing under the line profiles and assume a maximum error of 10\%. 

In contrast, the spectra acquired in the AR bright point (see Fig. \ref{Fig:spectra} and bottom panel of Fig. \ref{Fig:spectra_appendix} in the Appendix \ref{Sect:A2}) show broader profiles resulting in a worse fit and an increased error in the intensity estimation. In particular, it is not clear if the enhancements of the blue and red wings observed in some lines (such as the \siv~1406.93~\AA) might be due to a superposition of flows or to blending with other lines (such as \feii). In this case, we only fitted the main central component of the line profile and the difference between summed and fitted intensity is higher (Tab.\ref{tab:bright} in the Appendix \ref{Sect:A2} ). Therefore, we assume an error of 20\% for the intensity in the bright point spectrum. Similarly, a 20\% error is associated to the line intensity values in the flare case study (Sect. \ref{Sect:4.3}), where the line profiles are often asymmetric and need to be fitted with two or more Gaussian components. 
 
In Sect. \ref{Sect:4.2} and \ref{Sect:4.3} we present the analysis of the density diagnostics for the AR 12356 and the M 6.5 class flare on the 22 June 2015, respectively. We then discuss the results in Sect. \ref{Sect:5} and \ref{Sect:6}.

\begin{figure}[!ht]
	\centering
	\includegraphics[width=0.45\textwidth]{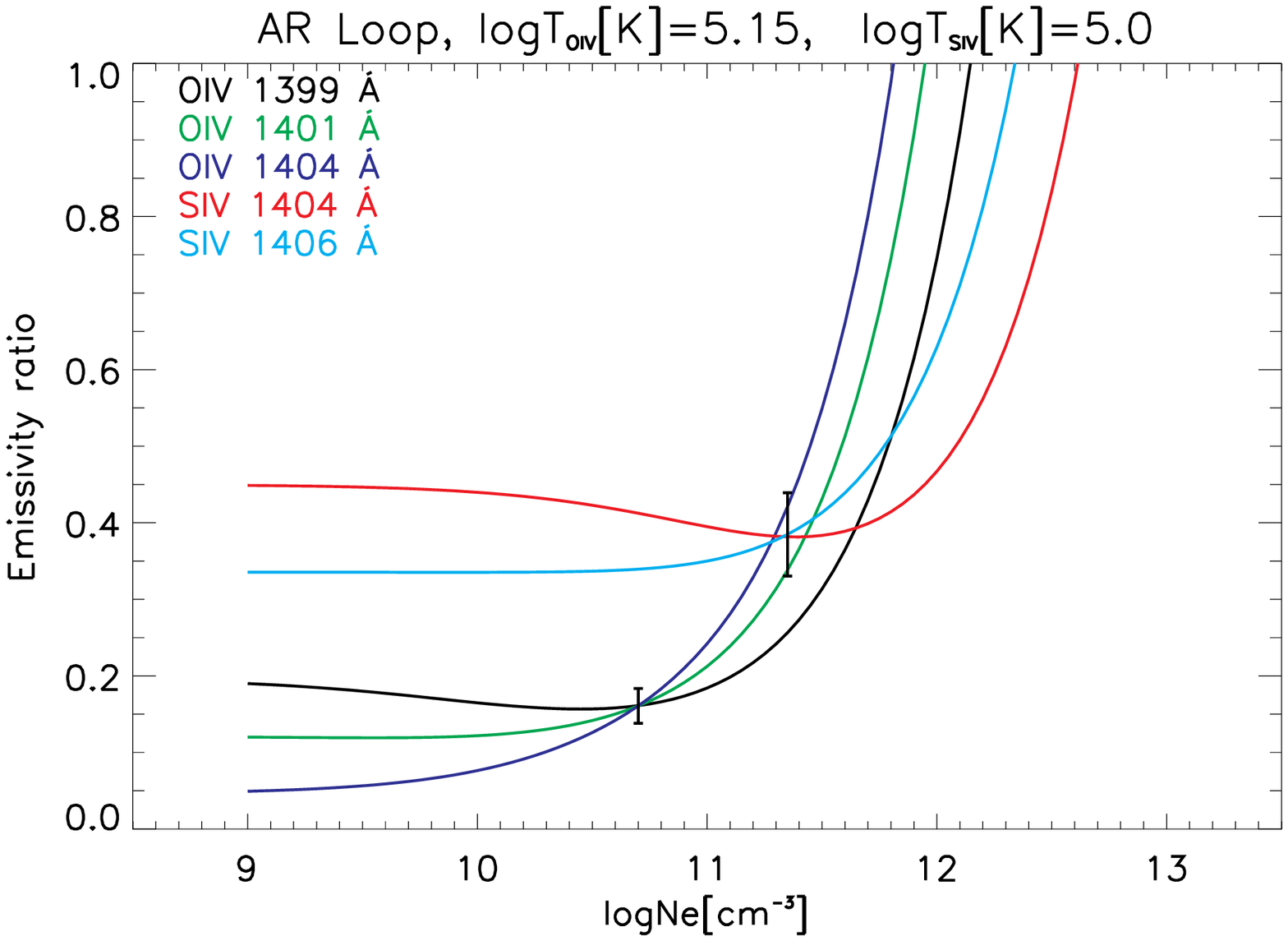} 	
	\includegraphics[width=0.45\textwidth]{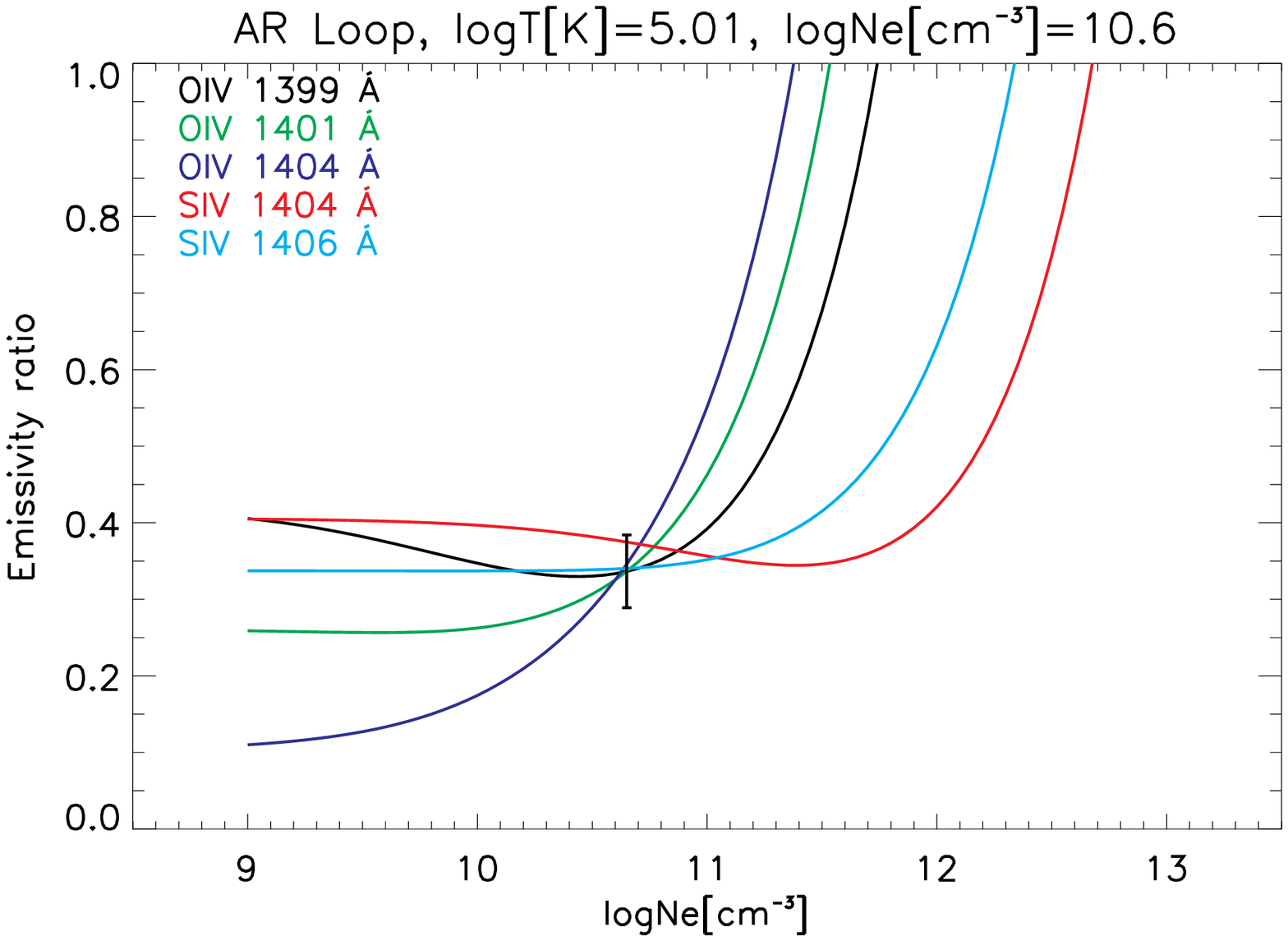} 
      \caption{Emissivity ratio curves as a function of density for the \oiv~and~\siv~spectral lines observed by IRIS in the "loop" region. Different colours for the curves indicate different spectral lines as described in the legend. In the top panel, we assume the typical temperatures of formation of log$T$[K] = 5.15 and 5 for the \oiv~and \siv~ions, respectively. In the bottom panel, a temperature of log$T$[K] $\approx$ 5 is assumed instead for both ions. The error bars include the propagation of a 10~\%~uncertainty in both the line intensity and the atomic data.}
      \label{Fig:EM_loci_loop}
\end{figure}
%
%
\begin{figure}[!hb]
	\centering
	\includegraphics[width=0.45\textwidth]{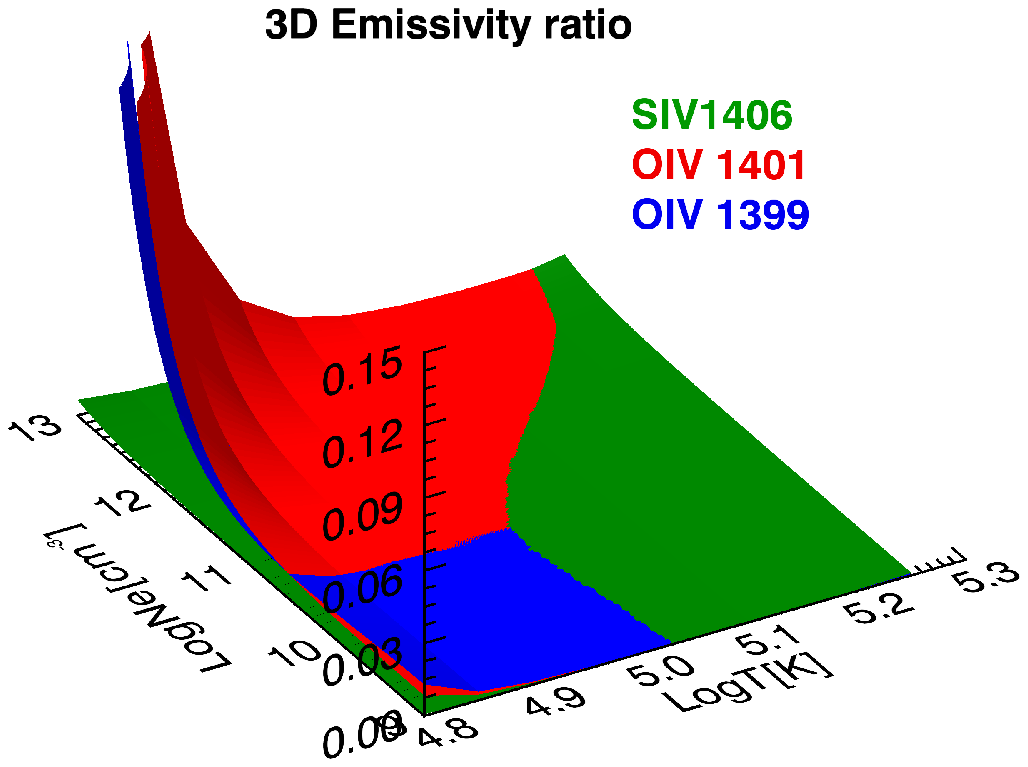} 	
      \caption{ 3D emissivity ratio as a function of log$N_\textrm{e}$[cm$^{-3}$] and log$T$[K] for the \siv~1406.93~\AA, \oiv~1399.78~\AA~and \oiv~1401.16~\AA~lines observed by IRIS in the "loop" region.  Different colours for the surfaces indicate different ions as described in the legend.}
      \label{Fig:surface_loop}
\end{figure}

\begin{table}
 \begin{center}
\caption{Line intensities $I_\textrm{obs}$ in the loop region}
 \begin{tabular}{llccc}
 \hline\hline\noalign{\smallskip}
 Ion & $\lambda$  &$I_\textrm{obs}$ (fit) & $I_\textrm{obs}$ (sum) & $\sigma$ \\
&\textbf{(\AA)}& \textbf{(*)}& \textbf{(*)}&\%\\
 \noalign{\smallskip}\hline\noalign{\smallskip}
 
 \siiv & 1402.77 &-&9758&1.0\\
 \oiv &1399.78  &541&533& 4.3\\
 \oiv & 1401.16 &1971&1958&2.3\\
  \siv & 1406.93 &451&449& 4.7 \\
  \oiv + \siv & 1404.82&561&561&4.2\\
%
\noalign{\smallskip}\hline
 \end{tabular}
 \label{tab:loop}
 \tablefoot{(*): $I_\textrm{obs}$ are expressed in phot s arcsec$^{2}$ cm$^{2}$. The intensities are obtained from the double-Gaussian fit (third column, see Fig. \ref{Fig:spectra_appendix}) and by summing the total counts (fourth row), as explained in the text. We take the error $\sigma$ associated to each $I_\textrm{obs}$ as either the Poisson standard deviation (taken as the square root of the total photon counts) or as the relative difference in the fitted and summed intensity values, when this difference is bigger than the Poisson error. The errors are expressed as a percentage of the $I_\textrm{obs}$ values.}
 \end{center}
 \end{table} 
   \begin{table*}
\centering
  \caption{Electron number density and temperature for different features in the AR 12356}
 \begin{tabular}{cccccccc}
 \hline\hline\noalign{\smallskip} 
AR Locations& \multicolumn{3}{c}{Log$N_\textrm{e}$ from ratios}& \multicolumn{2}{c}{Log$N_\textrm{e}$ from ER}&\multicolumn{2}{c}{Log$T$ from ER}\\
& $R_{1}$& $R_{3}$& $R_{4}$& $(A)$ &$(B)$& $(A)$ &$(B)$\\
& \multicolumn{3}{c}{(cm$^{-3}$)}& \multicolumn{2}{c}{(cm$^{-3}$)}&\multicolumn{2}{c}{(K)}\\
 \noalign{\smallskip}\hline\noalign{\smallskip} 
  Plage&  11.02 &  11.32& 12.68& 10.95&10.95&5.02&4.93\\
  Loop &10.70& 11.34 &12.17& 10.60&10.60&5.01&4.92\\
  BP &  11.06    & 11.08&12.49  &11.00&10.95&5.00&4.90\\
  \noalign{\smallskip}\hline
  \end{tabular}
 
  \tablefoot{Columns 2--4: electron number density for the three AR features obtained from the \oiv($R_{1}$), \siv($R_{3}$) and \siiv /\oiv($R_{4}$) line intensity ratios. $R_{1}$, $R_{3}$ and $R_{4}$ are calculated at the peak temperature of formation for each ion. Columns 5--8: density and temperature diagnostics from the $ER$ methods A and B respectively. The $ER$ method "$A$" uses ionization balances at low densities ($N_\textrm{e}$ $\approx$ 10$^{8}$ cm$^{-3}$). The $ER$ method "$B$" takes into account the $N_\textrm{e}$ dependence of the ionization balance, see Sect. \ref{Sect:5.1} for more detail.}
 \label{tab:densityAR}
 \end{table*} 

\begin{figure}
	\centering
	\includegraphics[width=0.45\textwidth, height=6cm]{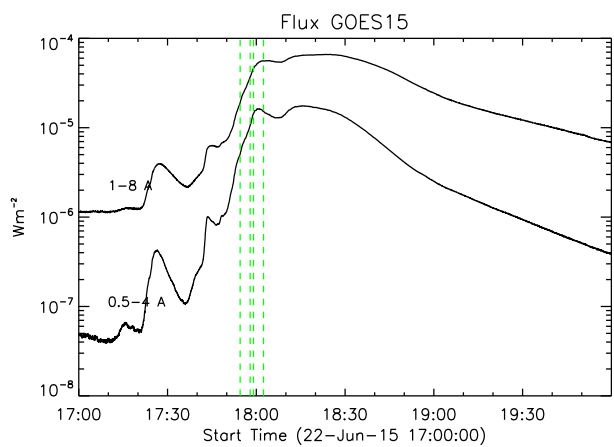} 	
      \caption{Soft X-ray light curves as measured by the GOES satellite in the 0.5-4~\AA~and 1-8~\AA~chennels during the 22-June-2015 flare. The dotted green lines represent the times of the IRIS rasters that were selected for measuring the electron density, as indicated in Table \ref{tab:density_flare}.}
      \label{Fig:goes}
\end{figure}
\begin{figure*}[!ht]
	\centering
	\includegraphics[width=\textwidth]{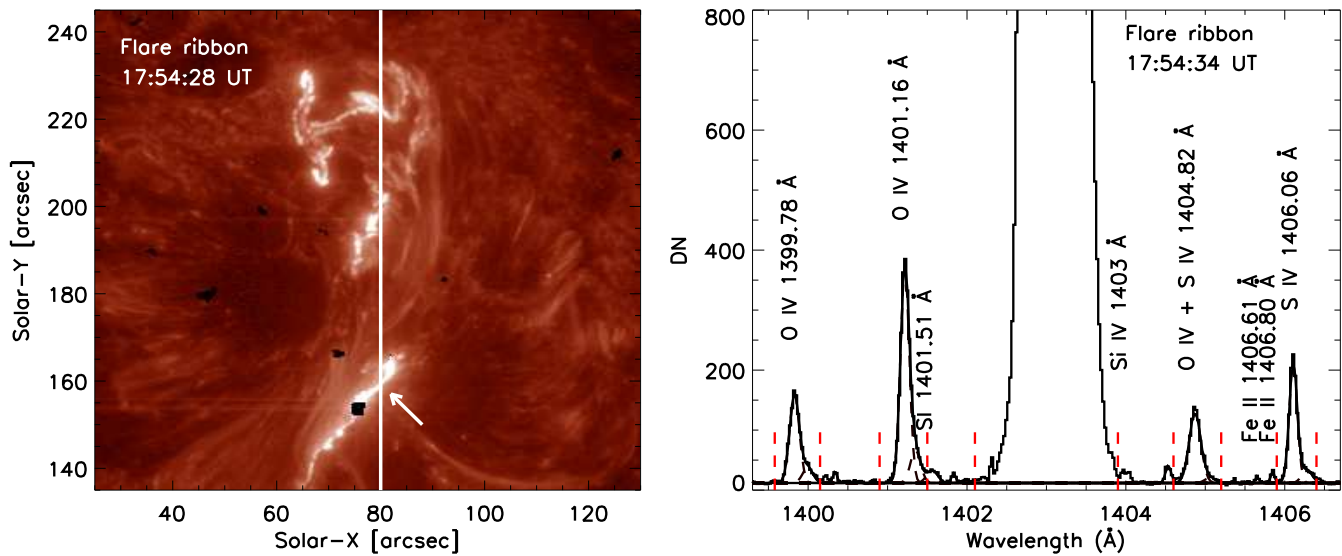} 	
      \caption{Left: IRIS Slit Jaw Image in the \siiv~1400~\AA~filter, showing the flare ribbons and the position of the IRIS spectrograph slit (vertical white line). The black pixels are due to dust contamination on the IRIS detector. Right: \siiv~window observed by the IRIS spectrograph at the southern flare ribbon, at the location indicated by the arrow. The dotted red lines represent the wavelength limits which were assumed when summing the total counts for each line. See text for further details (Sect. \ref{Sect:4.3}). }
      \label{Fig:SJI_spectrum}
\end{figure*}
\subsection{Density diagnostics in the AR NOAA 12356}
\label{Sect:4.2} 

The AR NOAA 12356 was visible during its passage on disk from May 28 to June 10, 2015 and during this time it did not show any major flaring activity. Fig. \ref{Fig:SJI} presents an overview of the AR 12356 as observed by IRIS in the SJI 1400~\AA~filter. This band is dominated by TR emission from \siiv~at around log$T$[K]=4.9 and shows several loop structures and compact bright points. Movie 1 shows the evolution of the AR over time, as observed in the SJI 1400~\AA~band. In particular, the movie shows that these loops and brightenings evolve very dynamically over time. The coloured box in Fig.\ref{Fig:SJI} indicates the field of view of the IRIS spectrograph monochromatic images shown in Fig. \ref{Fig:TR_maps}.
The IRIS spectrograph was running a single raster over the AR 12356 from 18:11 UT to 21:27 UT on the 1 June 2014. The dense 96-steps raster ran from east to west over a field of view of 33$\arcsec$ $\times$ 119$\arcsec$. Every slit position had a 0.33$\arcsec$ step size and an exposure time of $\approx$ 60 s. Several CCD spectral windows were included in this study, but we focus our analysis on the FUV spectral window centred around the \siiv\ 1402.77~\AA~line. 

The images in Fig. \ref{Fig:TR_maps} show the intensity maps of the \siiv\ 1402.77~\AA, \oiv~1399.78~\AA~and 1401.16~\AA, \siv~1406.06~\AA~spectral lines observed during the IRIS raster. For each slit exposure, the intensity at each detector pixel has been obtained by integrating the total counts (DN) over the line profiles and converting them to physical units (phot $\cdot$ s$^{-1}$ $\cdot$ arcsec$^{-2}$ $\cdot$ cm$^{-2}$). 

We analyse three different features in the AR 12356 where the \oiv\ and \siv\ lines were reasonably intense: a more steady region that we call "plage", a "loop structure" and a brightening that we call "bright point". These regions are indicated in Fig.\ref{Fig:TR_maps} by the yellow, light blue and red coloured boxes respectively. The corresponding IRIS TR spectra (black, blue and red lines) are shown in Fig. \ref{Fig:spectra}. The individual spectra are plotted in Fig. \ref{Fig:spectra_appendix} (Appendix \ref{Sect:A2}), which also shows the double Gaussian fit. We note that the line profiles are much broader in the bright point spectrum than in the loop and plage spectra, possibly due to a superposition of different flows along the line of sight or turbulent motion in the plasma. In addition, the relative intensity of the observed lines (in particular of the 1404.82~\AA~blend and the \siv\ 1406.06~\AA) changes significantly between the three spectra. This is a consequence of the density variation between different regions. Movie 1 shows these features over time at high spatial resolution with the IRIS SJI. In particular, the so-called "plage" region seems to be made by small unresolved arch-like structures.
 
In the following, we show detailed results for the loop region only and we refer the reader to the Appendix \ref{Sect:A2} for the AR plage and bright point analysis.


The top panel of Fig. \ref{Fig:EM_loci_loop} shows the emissivity ratios of each IRIS TR line observed in the loop spectrum as a function of density, given by eq. \ref{eq:ratio} and described in Sect. \ref{Sect:2}. The intensity of the blended \oiv\ 1404.81~\AA~and \siv\ 1404.85~\AA~lines was obtained as described in Eqs. \ref{eq:Ioiv} and \ref{eq:Isiv}. The temperature \textit{T} used to calculate \textit{G$_\mathrm{th}(T,N_\mathrm{e}$)} in Eq. \ref{eq:Ith} is initially taken to be the temperature of maximum abundance for each ion from the fractional ion abundances available in CHIANTI v.8; i.e., log$T$[K] $\approx$ 5.15 and 5.0 for \oiv\ and \siv\ respectively. It is important to note that these fractional ion abundances are calculated in CHIANTI assuming a low electron number density, i.e., around 10$^{8}$ cm$^{-3}$. As we mentioned in Sect. \ref{Sect:3}, given the small difference in temperature of formation of the two ions, if we assume pressure equilibrium, we expect the \oiv\ and \siv\ plasma to be formed at electron densities which differ by up to a factor of $\approx$ 1.8.

The two \siv\ and three \oiv\ transitions in the top panel of Fig. \ref{Fig:EM_loci_loop} show two distinct crossing points which would indicate electron densities log$N_\textrm{e}$ [cm$^{-3}$] of 10.70 and 11.34 respectively. The density values obtained by using the separate \oiv\ and \siv\ ratios thus differ of a factor by around 4, i.e., much higher than expected assuming constant pressure. The error bars at the crossing points represent the propagation of a 10\%~error in both the line intensities and the $G_\textrm{th}(T,N_\textrm{e})$. We emphasize that estimating the density from the crossing points of the \oiv\ and \siv\ lines separately is equivalent to applying the intensity ratio method (Eqs. \ref{eq:R1}--\ref{eq:R3}). We also note a significant discrepancy between the relative intensity of those ions taking into account the error bars. A discrepancy in the relative line intensity is also found in both the bright point and plage spectra (see Appendix \ref{Sect:A2}) as well as in the flare spectrum analysed in Sect. \ref{Sect:4.3}. We note that previous studies \citep[e.g., ][]{Keenan02} mainly focused on comparing density diagnostics using the \oiv\ and \siv\ ratios individually but did not report any comparison on the relative intensity of spectral lines from the two ions.

We rule out that the discrepancy in the relative intensity and derived electron density values could be due to an underestimation of the errors associated to either the  atomic data or observed intensities. In the Appendix \ref{Sect:A1} we show that atomic calculations from different authors agree better than a 10\%. Regarding the intensity values, we note that the intensity of the \oiv\ 1404.81~\AA~and \siv\ 1404.85~\AA~lines might indeed be affected by a larger error than the other unblended lines. However, the discrepancy between the unblended \oiv\ and \siv\ lines would still remain. In addition, we emphasize that possible blends of the \oiv\ and \siv\ with the photospheric \si\ and \feii\ spectral lines do not represent a major issue as discussed in Sect. \ref{Sect:4.1}. Finally, the discrepancy is observed in all AR features studied as well as in the flare spectrum presented in Sect. \ref{Sect:4.2}, suggesting that differences in O and S chemical abundances are unlikely to be the main cause. 

We recall that the emissivity ratio curves shown in the top panel of Fig. \ref{Fig:EM_loci_loop} are obtained assuming that the \oiv\ and \siv\ ions are formed at the temperature of maximum ionization abundance for each ion. We relaxed this assumption and investigated whether we could find a combination of a single density and a single temperature values for the emitting plasma which could reproduce the intensity of all the \oiv\ and \siv\ lines in the observed AR loop spectrum. Given the lack of suitable spectral lines to produce a DEM distribution, we start by assuming that the \oiv\ and \siv\ plasma are nearly isothermal. This approximation can be justified by the fact that AR loops are often observed to show a very narrow distribution in their cross-section \citep{delzanna:03, Warren08,Brooks12,Schmelz07}. This seem to be particularly true for cooler loops which tend to have narrower DEM distributions than hotter loops \citep{Schmelz14, Schmelz09}.
In Fig. \ref{Fig:surface_loop} we show a 3D plot of the emissivity ratio $ER(T,N_\mathrm{e})$ as a function of $T$ and $N_\mathrm{e}$ for just the three unblended lines observed by IRIS: \oiv\ 1339.77~\AA~and 1401.16~\AA, and \siv\ 1406.06~\AA. The 3D emissivity ratios consistently intersect at a temperature of log$T$[K] $\approx$ 5.0 and a density of log$N_{\textrm{e}}$ [cm$^{-3}$] $\approx$ 10.6. Hence, we recalculate the $G_\mathrm{th}(T, N_\textrm{e})$ function at this latter value of temperature for each of the observed IRIS TR lines. We note that the observed intensity $I_\mathrm{obs}$ of the blended \oiv\ 1404.81~\AA~line (and consequently of the \siv\ 1404.85~\AA~line) also changes slightly because it is now obtained from Eq. \ref{eq:Ioiv} by taking $R_{2}$ at the density log$N_{\textrm{e}}$ [cm$^{-3}$]  = 10.6 rather than at the density estimated by the $R_{1}$ ratio as previously. 

The bottom panel of Fig. \ref{Fig:EM_loci_loop} shows the new emissivity ratio curves obtained assuming the temperature and density values obtained from the 3D emissivity ratios in Fig. \ref{Fig:surface_loop}. We note that the discrepancy between the \oiv\ and \siv\ intensities is now removed and that all the lines, including the \oiv\ at 1404.81~\AA~ and the \siv\ at 1404.85~\AA, consistently indicate a near iso-thermal (log$T$[K] $\approx$ 5.0) and iso-density (log$N_{\textrm{e}}$ [cm$^{-3}$] $\approx$ 10.6) plasma. 

It is interesting to note that the analysis of the bright point and plage spectra (see plots in the Appendix \ref{Sect:A2}) are also consistent with the emitting plasma being isothermal and in all cases formed at temperatures log$T$[K] $\approx$ 5.0. 

In Tab. \ref{tab:densityAR} (middle row) we compare the estimated density $N_\textrm{e}$ for the loop region obtained from the intensity ratios $R_{1}$, $R_{3}$ and $R_{4}$ (columns 2--4) and the $ER$ method analysis presented above (column 5, indicated as "$A$"). Column 6 shows the density values obtained from the $ER$ method after taking into account the $N_\textrm{e}$ dependence in the fractional ion abundances of each ion (indicated as $ER$ method "$B$"). This latter method will be described in detail in Sect. \ref{Sect:5.1}. Finally, columns 7 and 8 show the estimated values of plasma temperature obtained by using the $ER$ method in the case $A$ and $B$ respectively. The same results for the plage and bright point regions are also reported in the top and bottom rows of Tab.\ref{tab:densityAR}. 

In Sect. \ref{Sect:4.3} we perform a similar density diagnostic analysis with spectra acquired during the impulsive phase of the 22 June 2015 flare. 
 
 
 
%

%

%

%
%
\begin{table*}
\centering
\caption{Electron number density and temperature at the flare ribbon over time}
 \begin{tabular}{cccccccc}
 \hline\hline\noalign{\smallskip} 
 Time  & \multicolumn{3}{c}{Log$N_\textrm{e}$ from ratios}& \multicolumn{2}{c}{Log$N_\textrm{e}$ from ER}&\multicolumn{2}{c}{Log$T$ from ER}\\
& $R_{1}$& $R_{3}$& $R_{4}$& $(A)$ &$(B)$& $(A)$ &$(B)$\\
(UT)& \multicolumn{3}{c}{(cm$^{-3}$)}& \multicolumn{2}{c}{(cm$^{-3}$)}&\multicolumn{2}{c}{K)}\\
 \noalign{\smallskip}\hline\noalign{\smallskip} 
  17:54:34 &  $>$ 12.00 & 13.1&13.64&&\\
  17:57:57 & $>$ 12.00& 13.22&-&&\\
  17:59:05 &  $>$ 12.00 & $\gtrsim$ 13.20&13.85&&\\
  18:02:28&   11.74&  11.96&13.43&11.70 & 11.65&4.98&4.87\\
   \noalign{\smallskip}\hline
  \end{tabular}
 \label{tab:density_flare}
  \tablefoot{Columns 2--4: electron number density at the flare ribbon over time obtained from the \oiv($R_{1}$), \siv($R_{3}$) and \siiv /\oiv($R_{4}$) line intensity ratios. $R_{1}$, $R_{3}$ and $R_{4}$ are calculated at the peak temperature of formation for each ion. Columns 5--8: density and temperature diagnostics from the $ER$ methods A and B respectively. The $ER$ method "$A$" uses ionization balances at low densities ($N_\textrm{e}$ $\approx$ 10$^{8}$ cm$^{-3}$). The $ER$ method "$B$" takes into account the $N_\textrm{e}$ dependence of the ionization balance, see Sect. \ref{Sect:5.1} for more detail.}
 \end{table*} 
%
%

%

\begin{figure}[!h]
	\centering
	\includegraphics[width=0.45\textwidth]{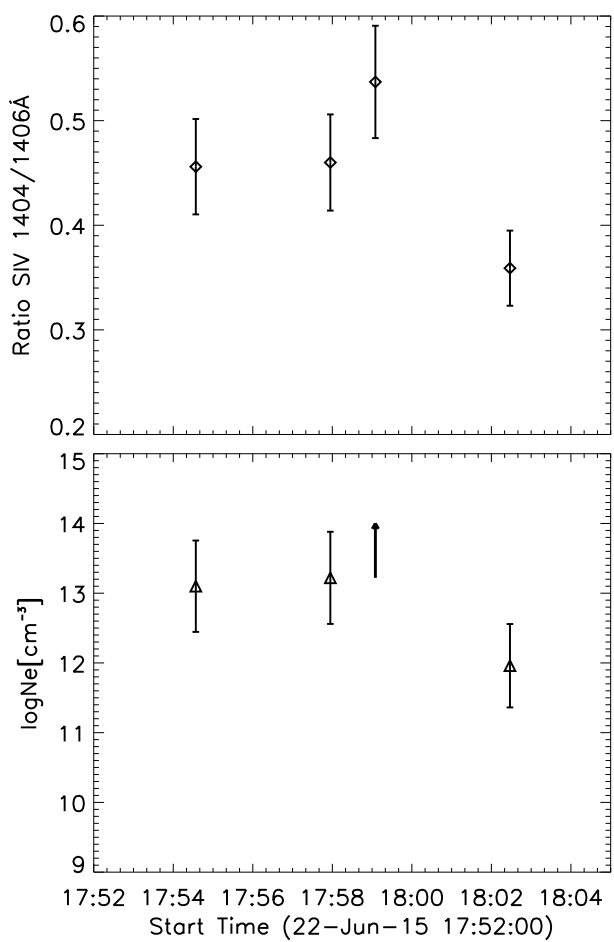} 	
      \caption{Ratio of \siv~1404/1406 lines (top panel) and derived electron density (bottom panel) as a function of time for the 22-June-2015 flare. The arrow indicates the high density limit ($\approx$ 10$^{13}$ cm$^{-3}$) of the \siv\ $R_3$ line ratio.}
      \label{Fig:density_time}
\end{figure}

\subsection{High density measurements during the 22 June 2015 M-class flare}   
\label{Sect:4.3}
In this section, we focus on the M6.5 class flare that occurred in AR NOAA 12371 on 22 June 2015. The flare started at about 17:23~UT and reached the maximum of the X-ray flux at around 18:00~UT, as shown in the GOES light curves in Fig. \ref{Fig:goes}. The peak phase was followed by a long gradual phase until about 20:53~UT.

On 22 June 2015, IRIS was running several 16-step sparse rasters observing the AR 12371 from 17:00~UT to 21:14~UT. Each raster had a FOV of 15$\arcsec$ $\times$ 119$\arcsec$ and a cadence of 33 s, with a single slit exposure time of just $\approx$ 2 s. 

The left panel of Fig. \ref{Fig:SJI_spectrum} shows that the IRIS slit (vertical line) intersected both northern and southern ribbons during the whole evolution of the flare. The right panel of Fig. \ref{Fig:SJI_spectrum} shows a spectrum of the \siiv\ 1402.77~\AA~window at the location where the slit was crossing the southern flare ribbon at around 17:54~UT, during raster No. 96 and at slit exposure No. 9. We note that the relative intensity of the \siv\ line at 1406.06~\AA~and the \oiv\ + \siv\ blend at 1404.82~\AA~is significantly larger compared to the AR spectra presented in Sect. \ref{Sect:4.2}. This is a result of the increase in the electron density. 

We analyse four TR spectra at the same slit position No.9, crossing the southern flare ribbon during the impulsive and peak phases of the flare, where the TR lines are strong. We only selected spectra where the line profiles were not strongly asymmetric or contaminated by a high FUV continuum. Table \ref{tab:density_flare} summarizes the results of the density diagnostics by using the intensity ratios $R_{1}$, $R_{3}$ and $R_{4}$ and emissivity ratio method "$A$" when applicable. In addition, the last two columns show the plasma density and temperature obtained by using the $ER$ method "$B$", i.e., taking into account the ionization balances in high density regime (with ADAS), as described in Sect. \ref{Sect:5.1}. 

The emissivity ratio methods ("$A$" or "$B$") cannot be applied if $R_\textrm{1}$ lies outside the density sensitivity limit of 10$^{12}$ cm$^{-3}$ (Fig. \ref{Fig:ratios} top). In such cases however, the \siv\ ratio can still be used as it is sensitive to densities up to 10$^{13}$ cm$^{-3}$ (Fig. \ref{Fig:ratios} bottom). It is important to note that the ratio $R_\textrm{2}$ also reaches a statistical equilibrium above 10$^{12}$ cm$^{-3}$ and it remains constant at the value of around 6. Therefore, we can use this value to estimate the contribution of the \oiv\ 1404.81~\AA~line to the blend at 1404.82~\AA~from Eq. \ref{eq:Ioiv} and thus obtain the intensity of the \siv\ line involved in the $R_\textrm{3}$ ratio (from Eq. \ref{eq:Isiv}). 

It is important to note that at these high densities, the intensity of the \oiv\ 1404.81~\AA~(and thus \siv\ 1404.85~\AA) line obtained from $R_\textrm{2}$ will not change with density as in the previous case of the AR loop presented in Sect. \ref{Sect:4.2}. The \siv\ ratio is therefore a more reliable density diagnostic in the high density regime (above 10$^{12}$ cm$^{-3}$). 

The values of \siv\ ratios and corresponding densities over time are shown in Tab. \ref{tab:density_flare} and in Fig. \ref{Fig:density_time} for the four selected spectra. Fig. \ref{Fig:density_time} shows that the density reaches a peak of above 10$^{13}$ cm$^{-3}$ during the impulsive phase of the flare at about 17:59~UT. It then drops dramatically by more than one order of magnitude by around 18:02~UT, i.e., 2 minutes after the peak of the flare in the soft X-ray flux. The fourth column of Tab. \ref{tab:density_flare} shows the density diagnostics obtained by using the \siiv / \oiv\ $R_4$ ratio for the four spectra analysed, except the one at ~17:57:57~UT where the \siiv\ line is saturated. Interestingly, the density diagnostics from the \siiv /\oiv\ and the \siv\ ($R_3$) ratios seem to show better agreement at two times (second and third row Tab. \ref{tab:density_flare}) during the impulsive phase of the flare compared to the lower density AR case (Sect. \ref{Sect:4.2}). However, we note that at those two times the intensity of the \siiv\ line is very large and only $\approx$ 5$\%$ below the saturation threshold, which might affect the intensity estimation. In addition, opacity effects may also cause a decrease in the \siiv\ intensity (see Section \ref{Sect:6}).

\begin{figure}
	\centering
	\includegraphics[width=0.45\textwidth]{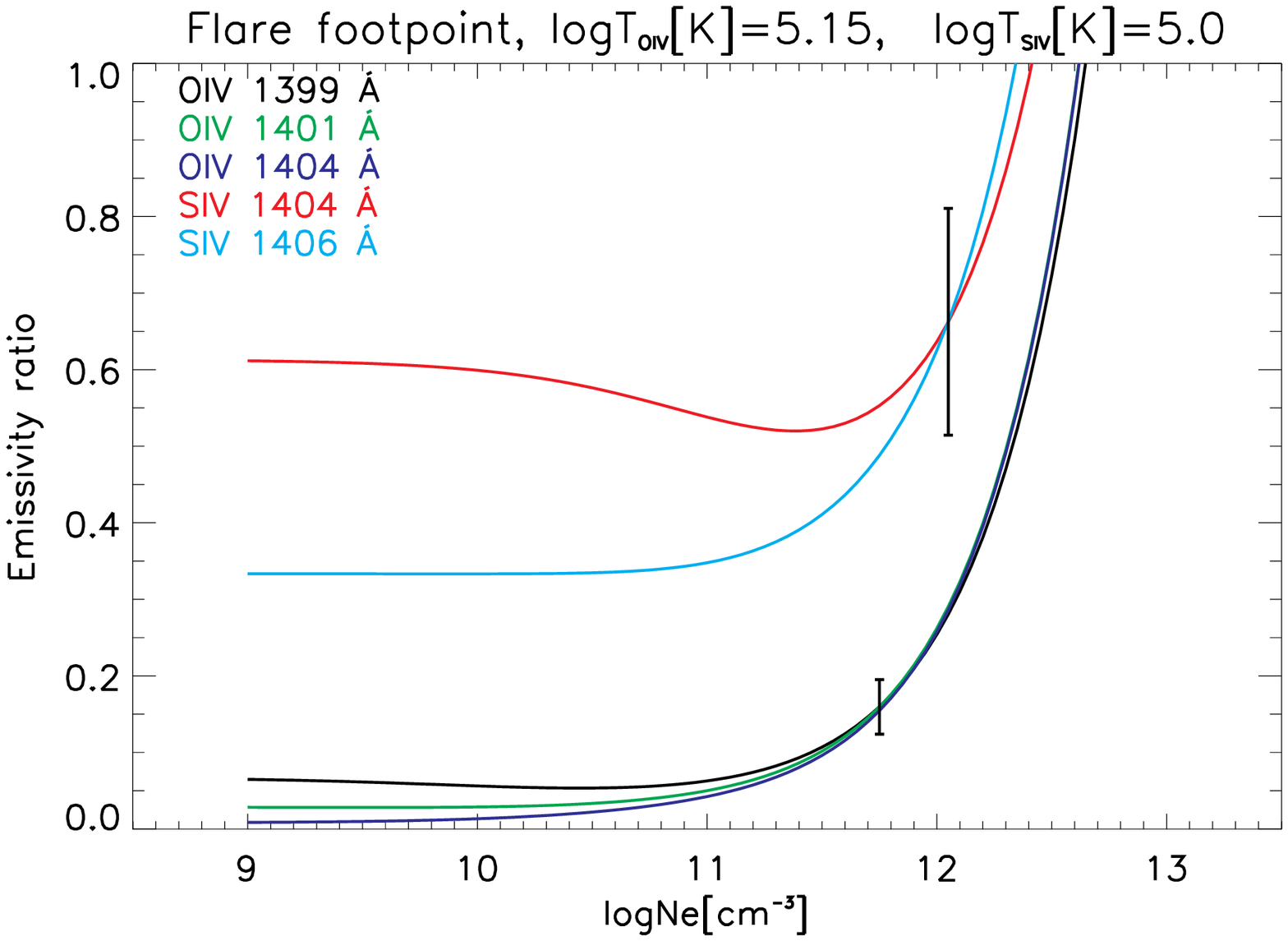} 	
	\includegraphics[width=0.45\textwidth]{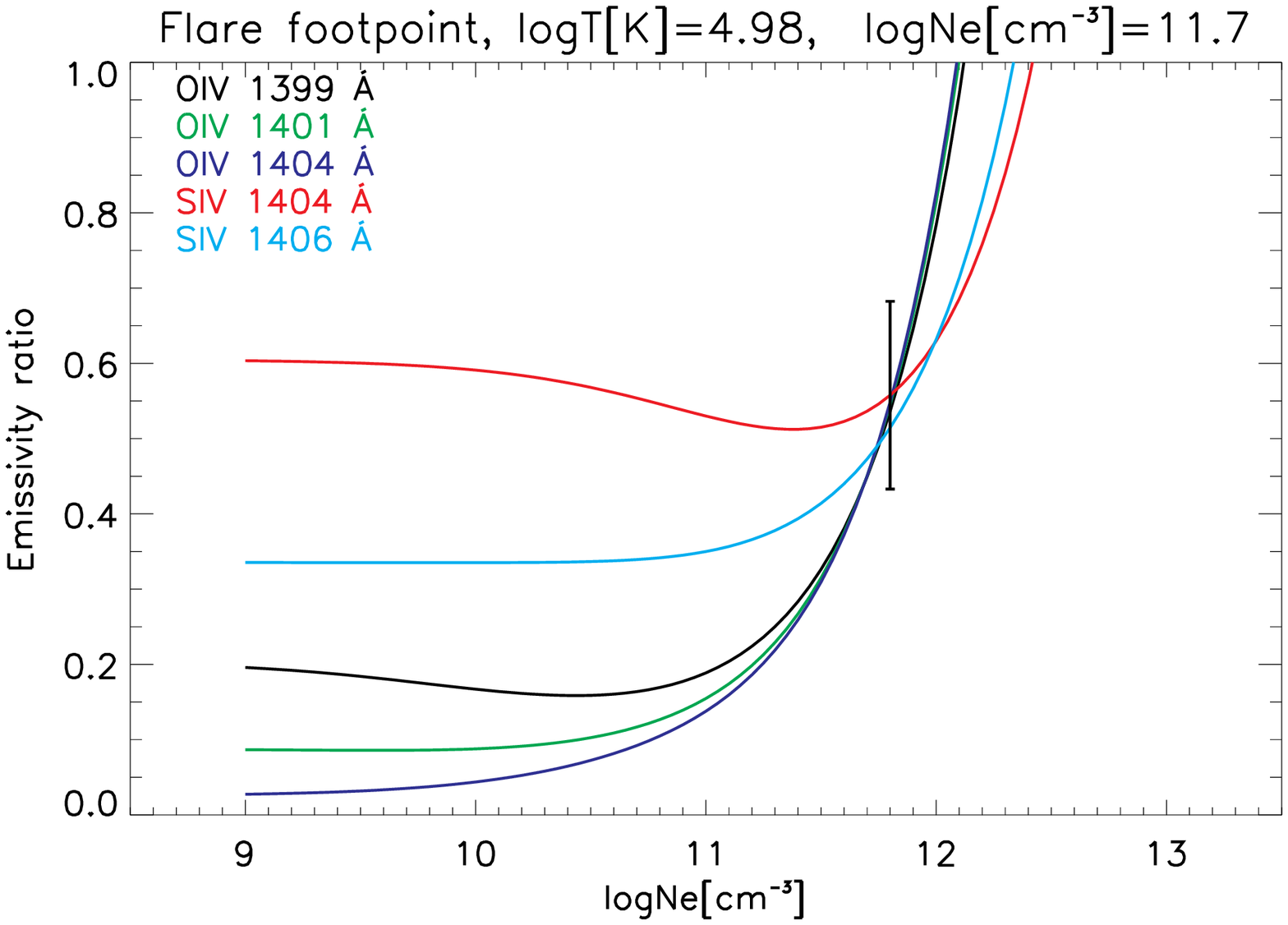} 	
      \caption{Emissivity ratio curves as a function of density for the \oiv~and~\siv~spectral lines observed by IRIS in the ribbon during the peak phase of the 22-June-2015 flare (at 18:02:28 UT). Different colours for the curves indicate different spectral lines as described in the legend. In the top panel, we assume the typical temperatures of formation of log$T$[K] = 5.15 and 5.0 for the \oiv~and \siv~ions, respectively. In the bottom panel, a temperature of log$T$[K] $\approx$ 4.98 is assumed instead for both ions. The error bars include the propagation of a 10~\%~uncertainty in both the line intensity and the atomic data.}
      \label{Fig:EM_loci_flare}
\end{figure}

\begin{figure}
	\centering
	\includegraphics[width=0.45\textwidth]{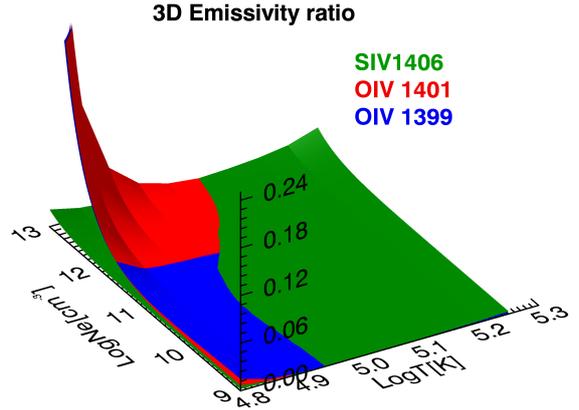} 	
      \caption{3D emissivity ratio as a function of log$N_\mathrm{e}$[cm$^{-3}$] and log$T$[K] for the \siv~1406.93~\AA, \oiv~1399.78~\AA~and \oiv~1401.16~\AA~lines observed by IRIS during the first peak of the 22-June-2015 flare ($\approx$18:02:28~UT, indicated by the last dotted vertical line in Fig. \ref{Fig:goes}).Different colours for the surfaces indicate different ions as described in the legend. }
      \label{Fig:surface_flare}
\end{figure}

 At 18:02\,UT, the density is again within the sensitive interval of the \oiv\ line ratios and we can thus apply the emissivity ratio method. The top panel of Fig. \ref{Fig:EM_loci_flare} shows the emissivity ratio curves for the \oiv\ and \siv\ lines observed by IRIS during the raster No. 110 at 18:02:28~UT. A very large discrepancy between the relative intensities of these lines is evident. Hence, we use an approach similar to the one presented in Sect. \ref{Sect:4.2} and try to determine whether the emitting plasma can be fitted with a different value of temperature, starting from a near-isothermal assumption for the two ions. Figure \ref{Fig:surface_flare} shows a 3D emissivity ratio plot. We find that the flare spectrum at 18:02\,UT is consistent with a plasma emission being isothermal at the temperature of log$T$[K] $\approx$ 4.98 and density log$N_\textrm{e}$ [cm$^{-3}$] $\approx$ 11.70. We then re-calculated the $G_\textrm{th}$ function at the newly estimated temperature and density, as explained in Sect. \ref{Sect:4.2}. The corresponding new emissivity curves are plotted in the bottom panel of Fig. \ref{Fig:EM_loci_flare}, showing that the intensity of all the observed \oiv\ and \siv\ lines, including also the \oiv\ 1404.81~\AA~and \siv\ 1404.85~\AA, are indeed consistent with a plasma being near iso-density and iso-thermal.

%
%
%

 
%
%
\section{Processes affecting the ion formation temperature}
\label{Sect:5}
Sect. \ref{Sect:4.1} and \ref{Sect:4.2} show that the observations of \oiv\ and \siv\ transition lines are consistent with an isothermal assumption and temperatures of about log$T$[K]\,$\approx$\,5.0 This value is consistent with the temperature of the \siv\ maximum ion abundance but it is lower than the temperature at which most of \oiv\ plasma is expected to emit in ionization equilibrium. 

As we already mentioned in Sect. \ref{Sect:4.2}, cooler coronal loops often show a near-isothermal distribution, so it is not surprising to see this result for the TR loops.
 It is not obvious to us why we should find an isothermal temperature for the \oiv\ and \siv\ plasma also at the flare ribbon, AR plage and bright point but this does indeed seem to fit the observed data well. As mentioned above, the online movie 1 shows that the so-called "plage" region is continuously filled with small and very fast loop-like structures. Hence, one possible explanation is that in all cases the plasma we observe either originates from loop structures (AR loop and possible the plage region) or loop footpoints (for the flare and bright point cases), which are both features where the plasma is often observed to be near isothermal.

However, the predicted formation temperature of the ions is uncertain. There are in fact several effects that can shift the temperature of formation of the ions, the most important ones being briefly summarised in Sect. \ref{Sect:5.1}, \ref{Sect:5.2} and \ref{Sect:5.3}.

\subsection{Density effects on the ion abundance}
\label{Sect:5.1}
The ion abundance can be obtained once the total effective ionisation and recombination rates are known. Most ion charge state distributions found in the literature are in the so-called low-density limit. Fractional ion abundances that take into account density effects were presented by \cite{burgess_summers:1969} and \cite{Jordan69}.
In particular, there are two main effects which become important at high densities, the suppression of the dielectronic recombination (DR), and the presence of metastable states.

As shown by \cite{burgess:1964}, DR is far more important that radiative recombination (RR). As first calculated in detail by \cite{burgess_summers:1969}, DR becomes suppressed at high densities. To derive appropriately the effects of finite density on the emissivity and so in particular on DR, complex Collisional-Radiative (CR) modelling needs to be carried out. Results of this modelling were published in a series of papers and reports by H. Summers \citep[see e.g.,][]{Summers74}. However, approximate cross-sections were used. For example, for the DR, the Burgess approximate formula was used. Since then, more accurate DR calculations in the low-density limit (the DR project, see \cite{badnell_etal:2003})
have been carried out. In addition, \cite{Nikolic13} have provided an approximate way to estimate the DR suppression by using the most recent DR rates and comparing the relative suppression with that which was calculated by H. Summers.

For several ions, significant population is present in the metastable states at high densities, hence 
ionisation and recombination from these states also needs to be taken into account. The role of metastable states and the effects of finite density on DR are considered within the generalised collisional-radiative (GCR) modelling in ADAS \citep{summers_etal:2006}. 

\begin{figure}[!ht]
	\centering
	\includegraphics[width=0.48\textwidth]{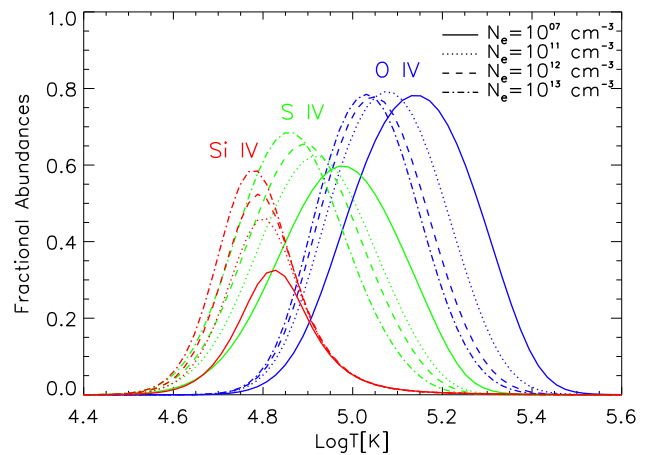} 	
      \caption{Fractional ion abundances $N(X^{+m})/ N(X)$ for \oiv\ (blue lines), \siv\ (green lines) and \siiv\ (red lines) calculated as a function of different electron number densities using atomic data from the OPEN-ADAS database. The continuous, dotted, dashed and dot-dashed lines indicate a different value of electron number density used to calculate the fractional ion abundances for each ion, as described in the legend. }
      \label{Fig:ioneq}
  \end{figure}

Fig. \ref{Fig:ioneq} shows the fractional ion abundances for \oiv, \siv\ and \siiv\ obtained by using the effective rate coefficients for ionisation and recombination available in the OPEN-ADAS database\footnotemark[1] (for \oiv, the GCR model was used; for \siv, the CR model was used). The continuous curves show the fractional ion abundances calculated at low density values ($N_\textrm{e}$ = 10$^{8}$ cm$^{-3}$) while the dotted lines show the results for higher densities ($N_\textrm{e}$ = 10$^{11-13}$ cm$^{-3}$). As already shown by \cite[e.g.,][]{vernazza_raymond:1979}, including the effects of DR suppression and the presence of metastable levels results in a shift the formation temperature of TR ions towards lower values. In a stratified atmosphere, lower temperature plasma regions have a larger density which thus cause an increase in the line intensities. This increase is particularly large for allowed lines, whose intensity is proportional to $N_\textrm{e}$ $^{2}$, and might explain the anomalously high intensity of some of such lines (such as the \siiv\ 1402.77~\AA). In addition, Fig. \ref{Fig:ioneq} shows that the values of the peak fractional ion abundances for \siiv\ and \siv\ increase significantly with density, contributing to an additional increase in the intensity of those ions.  

In order to estimate the effect on the density diagnostics, we applied the $ER$ method to the AR loop and flare observations by including the density-dependent ADAS fractional ion abundances in Eq. \ref{eq:Ith} .
The results of density and temperature so obtained are given in Tabs. \ref{tab:densityAR} and \ref{tab:density_flare} (indicated as $ER$ method "$B$"). In both cases, the estimated plasma temperature is lower than the temperatures that we obtained in Sect. \ref{Sect:4} using fractional ion abundances from CHIANTI (calculated at low density). In contrast, the density values do not change significantly, showing that the high density effects on the ion abundances do not affect the density diagnostic results presented in Sect. \ref{Sect:4}, in the isothermal approximation. 

\subsection{Non-equilibrium ionisation}
\label{Sect:5.2}
We note that IRIS does not observe spectral lines of successive stages of ionisation of an element which would enable us to study non-equilibrium ionisation in detail. The effect of transient ionization on the intensity of the \oiv\ and \siiv\ lines observed by IRIS at plasma densities of around 10$^{10}$ cm$^{-3}$ was recently investigated by \cite{Doyle13} and \cite{Olluri13}. Those lines show a very different response to transient ionization because of their different formation process. In contrast, the \siv\ and \oiv\ lines analysed in this work have similar atomic structure (see Tab. \ref{tab:iris_lines}) and therefore we expect them to behave in a similar manner under non-equilibrium ionization. \\
  
In addition, the relatively high densities we expect and measure in the low TR where \oiv\ is emitted in active regions ($\approx$ 10$^{11}$ cm$^{-3}$) and in the flare under study ($\approx$ 10$^{13}$ cm$^{-3}$), mean that strong departures from equilibrium ionisation are not expected \citep[c.f.,][]{Dzifcakova16}. At these densities, the typical ionisation/recombination timescales are in fact of the order of 1s \citep[e.g.,][]{Smith10}, i.e. comparable to or less than the IRIS exposure times. In particular, the exposure time of the AR observation presented in Sect. \ref{Sect:4.2} is around 60s, i.e., much longer than the expected ionization/recombination timescales. The non-equilibrium ionization could however still be present if there are advective flows carrying plasma from regions with different temperature \citep[e.g.,][]{Bradshaw03,Olluri13}. We however do not detect any strong Doppler shifts, apart from the systematic $\approx$ 5 km s$^{-1}$ redshift in the quiet Sun TR \citep[e.g.,][]{Brekke97}, in any of the observed \oiv\ and \siv\ spectra in the AR loop and flare studies (see for example Fig. \ref{Fig:spectra}). The \oiv\ and \siv\ plasma should therefore be close to the ionization equilibrium at the densities we observe in this work (above 10$^{10}$ cm$^{-3}$).

\begin{figure*}[!ht]
	\centering
	\includegraphics[width=0.48\textwidth]{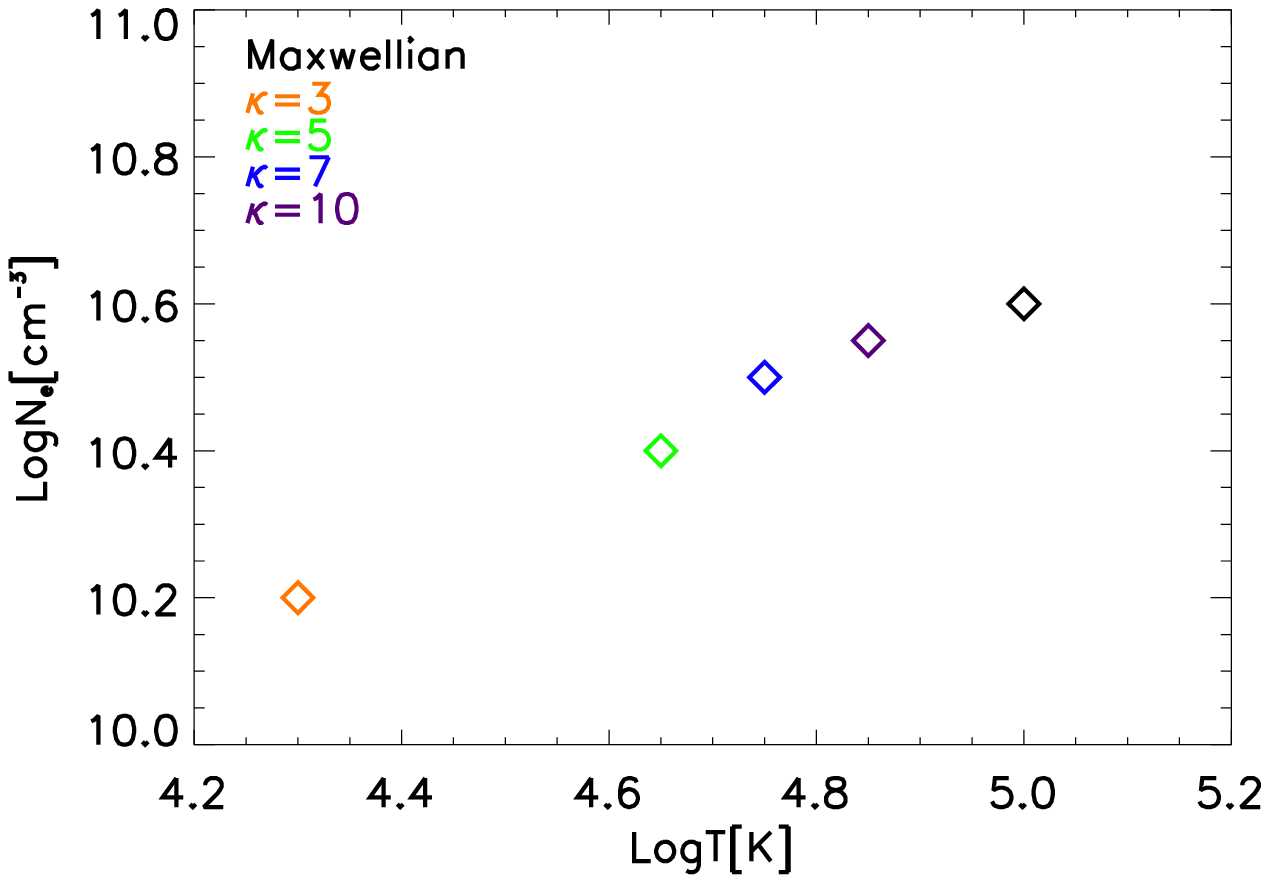} 	
	\includegraphics[width=0.45\textwidth]{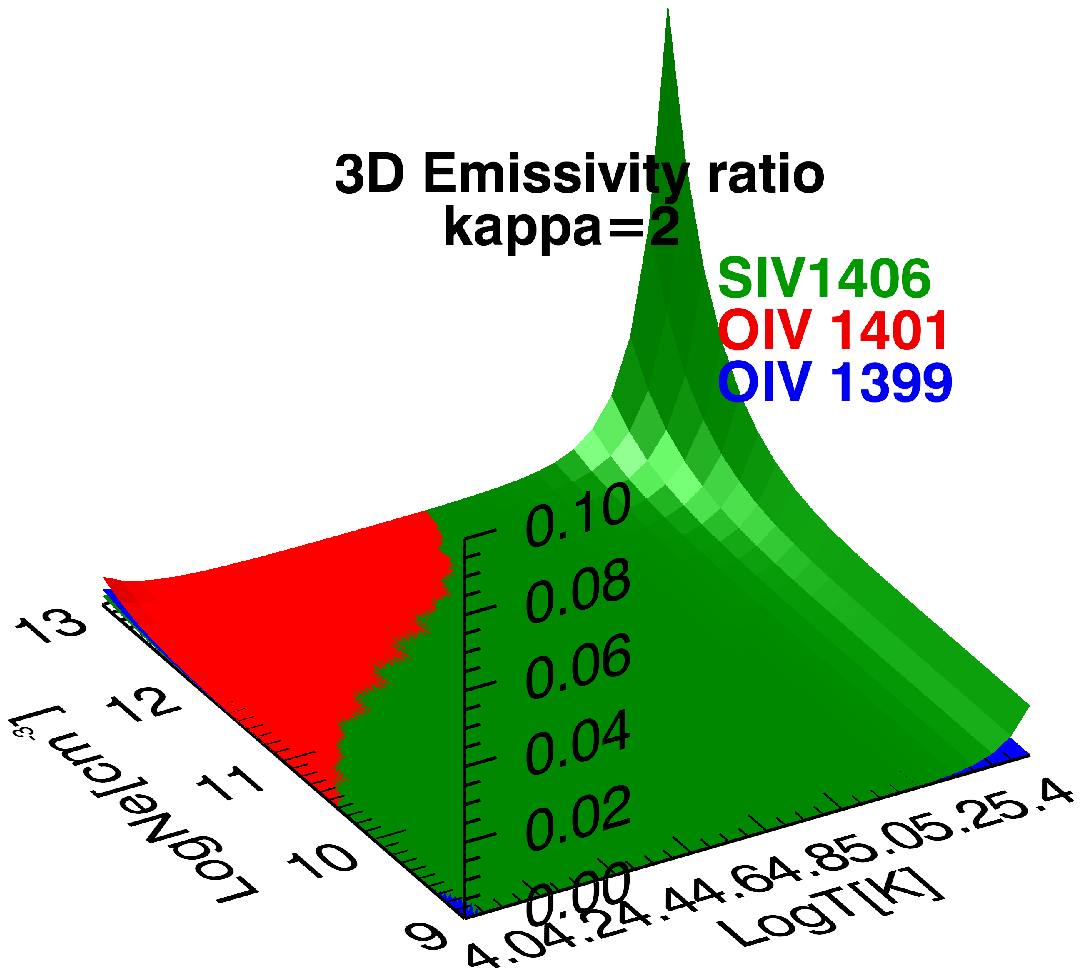} 	 
	
      \caption{Left: Density (x-axis) and temperature (y-axis) values at the 3D emissivity plot crossing points for the AR loop, obtained assuming non-Maxwellian distributions with different values of $\kappa$, except for $\kappa$ = 2. The density and temperature values for the Maxwellian case in Sect. \ref{Sect:4.2} are also shown for comparison. Right: 3D emissivity ratio for the $\kappa$ = 2 case. We note that there is no crossing point in this case. Different colours for the surfaces indicate different ions as described in the legend.}
      \label{Fig:k_values}
  \end{figure*}
%

%
%
%
%
%
%

\begin{table}
\centering
 \caption{Density and temperature estimation from the $ER$ method assuming non-Maxwellian electron distributions for different $\kappa$.}
 \begin{tabular}{ccc}
 \hline\hline\noalign{\smallskip} 
$\kappa$& Log$N_\textrm{e}$ from $ER$ ($A$)& Log$T$ from $ER$ ($A$)\\
 &  (cm$^{-3}$)& (K)\\
 \noalign{\smallskip}\hline\noalign{\smallskip} 
  2 & -&-\\
  3 &10.20 & 4.30\\
  5 & 10.40& 4.65\\
  7 &10.50 & 4.75 \\
  10 &10.55 &4.85 \\
   \noalign{\smallskip}\hline
\end{tabular}
\label{tab:densityAR_kappa}
\end{table}

\subsection{Non-thermal distributions}
\label{Sect:5.3}
 The presence of non-Maxwellian distributions with high-energy tails, such as the $\kappa$-distribution, could significantly shift the formation temperatures of the TR lines towards lower temperatures \citep{Dzifcakova11Si,Dzifcakova13}. For a DEM decreasing with temperature, the \siiv\ line will be strongly enhanced compared to the \oiv\ and \siv\ ones \citep{Dudik14}. This effect could easily explain the large intensities of the \siiv\ lines compared to the \oiv\ ones reported in Sect. \ref{Sect:4}, except for some of the spectra in the flare case (although see discussion in Sect.\ref{Sect:4.3}).

However, unambiguous detection of such non-Maxwellian distributions from IRIS observations is not trivial. This is due to the fact that only a few lines are observed by IRIS, and that these lines are formed from energy levels close in wavelength. There are however indirect indications of the possible presence of non-Maxwellian distributions in the present observations. First, as already mentioned, the \siiv\ intensities are too large by a significant factor compared to the Maxwellian predictions at the values of $T$ and $N_\mathrm{e}$ derived from the $ER$ method applied to the AR loop in Sect. \ref{Sect:4.2}, leading to the densities diagnosed from the ratio $R_4$ being much higher than using other methods (see Tab. \ref{tab:densityAR} in Sect. \ref{Sect:4.2}). Second, the line profiles observed in the AR (Sect. \ref{Sect:4.2}, Fig. \ref{Fig:spectra}), especially in the loop, show narrow cores and broad wings, consistent with a velocity distribution for the ions given by a $\kappa$-distribution (generalized Lorentzian). Preliminary work on fitting the line profiles using the method of \citet{Jeffrey16} suggests that at least some of the line profiles are well-fitted with a $\kappa$\,$\approx$\,2. Since a detailed analysis of multiple line profiles in different observed regions is beyond the scope of the present manuscript, in the following we only show the possible effect of the $\kappa$-distributions on the density diagnostics for the AR loop presented in Sect. \ref{Sect:4.2}. This can be done by calculating the contribution function $G_\textrm{th}$ of the \oiv\ and \siv\ lines assuming $\kappa$-distributions for the electrons, analogous to \citet{Dudik14}, but with the same atomic data as described in Sect. \ref{Sect:2}. Additionally, we have corrected a minor error in the collision strengths calculations performed by \citet{Dudik14} as shown in Appendix \ref{Sect:A3}. The estimated values of temperature and density from the $ER$ method for different $\kappa$ values are shown in the left panel of Fig. \ref{Fig:k_values}. The isothermal crossing points exist at for $\kappa$\,$\geqq$\,3 if temperatures of log$T$[K]\,$\geqq$\,4 are considered. The isothermal and iso-density crossing points occur at progressively lower $T$ and $N_\mathrm{e}$ (Fig. \ref{Fig:k_values}, Table \ref{tab:densityAR_kappa}). However, no crossing point exists for $\kappa$\,=\,2 (right panel of Fig. \ref{Fig:k_values}), which is inconsistent with the line profiles, assuming that the electrons and ions have the same velocity distribution. The cause of this is unknown at present. 

Tab. \ref{tab:densityAR_kappa} and Fig. \ref{Fig:k_values} show that a range of possible values of density and temperature diagnostics for the AR loop if non-thermal distributions at different $\kappa$ values were present in the observed plasma.


\section{Discussion and Summary}
\label{Sect:6}
In this work we have investigated the use of the \oiv\ and \siv\ emission lines near 1400~\AA~observed by IRIS as electron density diagnostics in the plasma from which they are emitted. These ions are formed at similar temperatures and therefore are expected to provide similar density diagnostics (within a factor of around two). 

Density diagnostics are usually based on the use of the intensity ratio of two lines from the same ion. We have applied an emissivity ratio method to obtain a value of electron density which can reproduce the relative intensity of the all the \oiv\ and \siv\ lines in the observed spectra. 


In our analysis, we have selected different plasma regions within an AR (loop, bright point and a plage region) and at the ribbon of the 22 June 2015 flare, where the lines were observed to be more intense. The \oiv\ and \siv\ lines (in particular the \oiv\ line at 1399.77~\AA) are usually very weak and cannot easily be detected in quiet Sun regions with IRIS.  

In all the features we analysed, we find that the \oiv\ and \siv\ lines give consistent density diagnostic results when we assume that the plasma has a near isothermal distribution rather than different temperatures of formation for the two ions. In all cases, the results are consistent with the plasma being at a temperature of about log$T$[K]=5. This temperature is lower than the peak formation temperature of \oiv\ calculated in CHIANTI v.8 assuming ionization equilibrium. However, a significant amount of \oiv\ is still formed at a temperature of log$T$[K]\,$\approx$\,5.0, as shown in the fractional abundance plot in Fig. \ref{Fig:ioneq} (continuous lines), and even more so when the high density effects are taken into account (dotted lines). The hypothesis of iso-thermality for the \oiv\ and \siv\ plasma could be explained by the fact that all the features we analysed are either cool loop structures or loop footpoints, which are often observed to be dominated by plasma with a very narrow thermal distribution \citep[e.g.][]{delzanna:03, Warren08, Schmelz07,Schmelz14}. We also emphasize that we are estimating average values of temperature and density for the emitting plasma.

Using the emissivity ratio method, we find electron number densities ranging from log$N_\textrm{e}$ [cm$^{-3}$] $\approx$ 10.6--11.0 in the AR loop and bright point respectively. The density variation in different plasma features in the AR under study can also qualitatively be estimated by comparing the spectra shown in the right panel of Fig. \ref{Fig:spectra}. For instance, in the bright point spectrum (red), formed in a higher density plasma, the intensity of the \siv\ 1406.1~\AA~spectral line is enhanced compared to the intensity of \oiv+\siv~blend at 1404.82~\AA. In contrast, in the loop spectrum the \siv\ 1406.16~\AA~line is weaker, indicating that the plasma is at a lower density. In addition, we note that the densities obtained by using the \siiv /\oiv\ line ratio $R_{4}$ are much higher than the values obtained by using the \oiv\ $R_{1}$ and \siv\ $R_3$ ratios, as shown in Tab. \ref{tab:densityAR} and previously noted by \cite{hayes_shine:1987}. This could be due to a number of issues, as outlined in Sect. \ref{Sect:1} and \ref{Sect:5.2}, and in particular to the anomalous behaviour of the Na-like ions. 

In the flare case presented in Sect. \ref{Sect:4.2}, the \oiv\ line ratio indicates a very high electron number density above the high density limit of $\approx$ 10$^{12}$ cm$^{-3}$.  The \siv\ line ratio is sensitive to higher electron densities and has been used to measure densities of $\approx$ 10$^{13}$ cm$^{-3}$ at the TR footpoints of the flare during the impulsive phase. Indications of high electron densities in the TR plasma during flares have been reported by some authors in the past. In particular, \cite{Keenan95} obtained 10$^{12}$ cm$^{-3}$ using the \ov\ line ratio, while \cite{Cook94} reported density values of 10$^{12.6}$ cm$^{-3}$ using ratios of allowed and intersystem \oiv\ lines. However, most of the other studies \citep[e.g.,][]{Cook95} were based on the use of line ratios, such the \oiv\ ones used in this work, which are density-sensitive up to a high density limit of 10$^{12}$ cm$^{-3}$, and therefore could only provide lower estimates for the electron density. Other authors have investigated electron number densities in high temperature plasma during flares. For instance, \cite{Doschek81} measured densities of 10$^{12}$~cm$^{-3}$ in the \ovii\ coronal plasma at 2 MK. Of particular importance is the study of \cite{Phillips96}, who showed for the first time very high electron densities (up to 10$^{13}$ cm$^{-3}$) from ions formed at $\approx$ 10~MK. Those high densities were observed 1 minute after the peak of a M-class flare. 
By using the ratio of the \siv\ lines observed by IRIS, we obtain an accurate electron density estimates for the TR plasma which are almost everywhere within the sensitivity range of the line ratio. Fig. \ref{Fig:density_time} shows that very high densities close to or above 10$^{13}$ cm$^{-3}$ are only reached over a short period of time during the peak of the flare, before dropping dramatically by more than an order of magnitude at the same footpoint position over 2 minutes. To the best of our knowledge, this is the first time we can directly diagnose such high electron number densities in the TR plasma during a flare. Density diagnostics based on the use of \siv\ line ratios with previous instruments in the past were complicated by the presence of line blends which could not be properly resolved and problems in the atomic data \citep{Dufton82,Cook95}, as also pointed out by \cite{keenan_etal:2002}. In this work we have shown that the \siv\ 1404.85~\AA~line can be accurately de-blended from the \oiv\ 1404.85~\AA~line in the high density interval above 10$^{12}$ cm$^{-3}$. In this case in fact the ratio $R_2$ remains constant, reducing the uncertainty associated with the estimation of the \oiv\ 1404.85~\AA~contribution to blend by using the \oiv\ 1401.16~\AA~line intensity (see middle panel of Fig. \ref{Fig:ratios}). 
One might think that at such high values of densities, the opacity effects may become important. This is not the case for the \oiv\ and \siv\ intercombination lines, due to the low $A$-values of these transitions. The optical depth can in fact be easily estimated by using the classical formula given for instance in \cite{Buchlin09}. We found that at a density of $\approx$~10$^{13}$ cm$^{-3}$, the \oiv\ lines reach an opacity of 1 over an emitting layer of the order of 10$^{5}$ km, which is much higher than the source size ($\approx$ size of the IRIS pixel, i.e. around 200 km). In contrast, the \siiv\ reaches an opacity of 1 over a  considerably smaller layer, of the order of $\approx$~20~km. This implies that opacity effects might be important for this line in the flare case study. In particular, a decrease in the \siiv\ intensity due to the opacity might result in wrong density estimates based on the \siiv /\oiv\ ratio.

Moreover, we emphasize the importance of including the effect of high electron number densities (above 10$^{10}$ cm$^{-3}$) in the calculations of the fractional ion abundances. 
 In Sect. \ref{Sect:5.1} we show that including fractional ion abundances for \oiv\ and \siv\ calculated at higher electron number density does not significantly affect the results of the density diagnostics from the emissivity ratio method. In contrast, the temperature formation of the ions, and therefore the temperature estimation from the emissivity ratio method, is shifted to lower values, as shown in Fig. \ref{Fig:ioneq}.

 Similarly, the presence of non-Maxwellian electron distribution in the plasma causes a shift of the temperature of formation of the ions to lower values. It is not possible to unambiguously detect signatures of non-thermal plasma conditions in the present study. A possible signature might arise from the analysis of the spectral line profiles. These profiles provide information regarding the velocity distribution of the ions, and it is reasonable to assume that this distribution is the same than the electron velocity distribution at these plasma densities and at the timescale of our observations. However, this analysis is quite involved and requires further investigation which is beyond the scope of this paper.  In this work, we are interested in estimating how the possible presence of non-thermal electron distributions might affect our density diagnostics. We therefore provide a range of possible values of density and temperature diagnostics for the AR loop obtained by using the emissivity ratio method, assuming that $\kappa$-distributions at different $\kappa$ values were present. The results  are presented in Fig. \ref{Fig:k_values} and Tab. \ref{tab:densityAR_kappa}, showing that the  density and temperature diagnostics would indeed differ significantly from the values obtained in Sect. \ref{Sect:4.2}, where we assumed Maxwellian electron distributions.\\

We have shown that combining \oiv\ and \siv\ observations from the recent IRIS satellite provide a useful tool to measure the electron number density in a variety of plasma environments. In particular, thanks to very high spatial resolution of IRIS ($\approx$ 200 km) it is now possible to select small spatial elements, reducing the problem of observing emission from very different density and temperature plasma regions, as pointed out in the past by e.g., \cite{Doschek84}. \\

In this work, we greatly emphasized the importance of including the \siv\ lines in the IRIS observational studies. Simultaneous, high-cadence observations of several spectral lines formed in the transition region can be used as direct density and temperature diagnostics of the emitting plasma. These diagnostics provide crucial information which can be compared with the predictions made by theoretical models of energetic events in the solar atmosphere. 
 
\begin{acknowledgements}
We thank the referee for the useful suggestions which helped improving the manuscript.
VP acknowledges support from the Isaac Newton Studentship, the Cambridge Trust, the IRIS team at Harvard-Smithsonian Centre for Astrophysics and the RS Newton Alumni Programme. GDZ and HEM acknowledge support from the STFC and the RS Newton Alumni Programme. JD acknowledges support from the RS Newton Alumni Programme. JD also acknowledges support from the Grant No. P209/12/1652 of the Grant Agency of the Czech Republic. AG acknowledges the in house research support provided by the Science and Technology Facilities Council. KR is supported by contract 8100002705 from Lockheed-Martin to SAO.

 IRIS is a NASA small explorer mission developed and operated by LMSAL with mission operations executed at NASA Ames Research center and major contributions to downlink communications funded by the Norwegian Space Center (NSC, Norway) through an ESA PRODEX contract. AIA data are courtesy of NASA/SDO and the respective science teams. CHIANTI is a collaborative project involving researchers at the universities of Cambridge (UK), George Mason and Michigan (USA). ADAS is a project managed at the University of Strathclyde (UK) and funded through memberships universities and astrophysics and fusion laboratories in Europe and worldwide.

\end{acknowledgements}

%
%

\begin{appendix}
\section{Atomic data and wavelengths}
\label{Sect:A1}

\begin{table*}[!htbp]
\begin{center}
\caption{A-values (s$^{-1}$) of the main \ion{O}{iv} transitions within 2s$^2$ 2p and 2s 2p$^2$. }
\begin{tabular}{llllllll}
 \hline\hline\noalign{\smallskip} 
$i-j$ &  Transition &  $\lambda$  (\AA)  & CH04  & TF02 & L12  &  R12 & \\
 &   &  (\AA)  &(s$^{-1}$)  &(s$^{-1}$)  &(s$^{-1}$)   & (s$^{-1}$)  & \\
 \noalign{\smallskip}\hline\noalign{\smallskip} 

1-2 & $^2$P$_{1/2}$ -- $^2$P$_{3/2}$ & 258933.  & 5.22$\times$10$^{-4}$ & 5.22$\times$10$^{-4}$ & 5.17$\times$10$^{-4}$ & 5.31$\times$10$^{-4}$ & \\

1-3 & $^2$P$_{1/2}$ -- $^4$P$_{1/2}$ &  1399.776 & 1.47$\times$10$^{3}$  &1.49$\times$10$^{3}$ & 1.79$\times$10$^{3}$& 1.46$\times$10$^{3}$ & \\
2-3 & $^2$P$_{3/2}$ -- $^4$P$_{1/2}$ &  1407.384 & 1.45$\times$10$^{3}$ & 1.47$\times$10$^{3}$ & 1.89$\times$10$^{3}$ & 1.44$\times$10$^{3}$ &  \\

1-4 & $^2$P$_{1/2}$ - $^4$P$_{3/2}$ &  1397.226 & 3.75$\times$10$^{1}$ & 3.91$\times$10$^{1}$ & 5.23$\times$10$^{1}$& 4.07$\times$10$^{1}$ & \\
2-4 & $^2$P$_{3/2}$ - $^4$P$_{3/2}$ &  1404.806 & 2.89$\times$10$^{2}$ & 2.94$\times$10$^{2}$ & 3.57$\times$10$^{2}$ & 2.85$\times$10$^{2}$ &  (bl) \\

2-5 & $^2$P$_{3/2}$ - $^4$P$_{5/2}$ &  1401.163 &1.16$\times$10$^{3}$ & 1.19$\times$10$^{3}$ & 1.24$\times$10$^{3}$ & 1.19$\times$10$^{3}$ &    \\
\noalign{\smallskip}\hline
\end{tabular}
\normalsize
\tablefoot{CH04: \cite{correge_hibbert:2004}; TF02: \cite{tachiev_froese_fisher:2000};  
 L12: \cite{liang_etal:2012}; R12: \cite{rynkun_etal:2012}. }
\end{center}
\label{tab:ne_o_4}
\end{table*}

\begin{table*}[!htbp]
\begin{center}
\caption{Wavelengths (\AA) of the \ion{O}{iv} IRIS  2s$^2$ 2p -- 2s 2p$^2$ transitions. }
\begin{tabular}{llllllll}
 \hline\hline\noalign{\smallskip} 
  Transition &   B69       & E83     & S86      &  Y11  & CHIANTI V.8 & NIST & \\
             &              &         &         & (CHIANTI v.7.1) &  & & \\
             &     (\AA)  & (\AA) & (\AA)& (\AA) & (\AA)  &(\AA) \\
 \noalign{\smallskip}\hline\noalign{\smallskip} 


$^2$P$_{1/2}$ -- $^4$P$_{1/2}$ & 1399.774 ($<$0.01) & 1399.776 & 1399.775 (.005) & 1399.766 (0.03) & 1399.776 &  - & \\
$^2$P$_{3/2}$ -- $^4$P$_{1/2}$ & 1407.386 ($<$0.01) & 1407.380 & 1407.386 (.005) & 1407.372 (0.03) & 1407.384 &  1407.382 & \\

$^2$P$_{1/2}$ - $^4$P$_{3/2}$ &  1397.20  (bl)   &  1397.214 & 1397.221 (0.005) & 1397.199 (0.03) &  1397.226 & - & \\
$^2$P$_{3/2}$ - $^4$P$_{3/2}$ &  1404.812 ($<$0.01) &  1404.790 & 1404.82 (0.01)  & 1404.783 (0.03) &  1404.806 & - & (bl) \\

$^2$P$_{3/2}$ - $^4$P$_{5/2}$ &  1401.156 ($<$0.01) &  1401.169 &  1401.163  (.005) & 1401.157 (0.03) & 1401.163 &  1401.157 & \\
\noalign{\smallskip}\hline
\end{tabular}
\normalsize
\tablefoot{Wavelengths are in vacuum. 
B69: \cite{bromander:1969}; E83: \cite{edlen:1983}; S86: \cite{sandlin_etal:1986}; 
Y11: \cite{young_etal:2011} and CHIANTI v.7.1; V.8: CHIANTI v.8 \citep{delzanna_etal:2015_chianti_v8}.
Values in brackets indicate the uncertainties in the measured wavelengths. }
\label{tab:ne_o_4b}
\end{center}
\end{table*}

\subsection{\siv}
In this section we discuss in detail the atomic data used for \siv.

Large discrepancies between the densities obtained from the \oiv~and \siv~lines have been noted by several authors. For example,  \cite{Cook95} presented solar and stellar observations obtained with the HRTS, the SO82B spectrograph on board \textit{Skylab} and the GHRS on board the HST.
The \oiv\ intensities were in the high-density limit during solar flares, and the \siv~ratios were inconsistent with the available atomic data. \cite{Cook95} used the \cite{dufton_etal:1982} scattering calculations for \siv. The main problem turned out to be that these atomic data for \ion{S}{iv} were incorrect. 
\cite{tayal:2000} carried out a new $R$-matrix scattering calculation that included 52 fine-structure levels. The new atomic data resolved the main discrepancies at low densities, as shown e.g. 
by \cite{keenan_etal:2002} using RR Tel HST STIS observations. 

However, the high-density solar and stellar observations as reviewed by \cite{DelZanna02} still indicated a small discrepancy in the 1404.8~\AA\ blend. Assuming that the \oiv~and \siv~ lines were formed at the temperatures near peak abundance in equilibrium, the densities obtained from all the \oiv~and \siv~ lines were consistent, with the exception of the 1404.8~\AA\ line, which was about 30\% stronger than predicted. 

These discrepancies prompted a new much larger $R$-matrix scattering calculation by the UK APAP  network\footnote{www.apap-network.org} and a revision of the wavelengths for this ion, presented in 
\cite{delzanna_badnell:2016}. We use this new atomic data here. Problems  were found in some of the atomic data calculated by \cite{tayal:2000}. However, they did not significantly affect the intensities of the intercombination lines, even though 715 fine-structure levels were included in the calculation. As shown in \cite{delzanna_badnell:2016}, the main excitation rates that drive the level population of the 3s 3p$^{2}$ $^{4}$P levels are very close (to within a few percent) to those calculated by \cite{tayal:2000}.

A Table of A-values calculated by various authors was also provided in \cite{delzanna_badnell:2016}.
The calculated A-values for the main transitions were within a few percent of the intensities of the most accurate previous calculations, by \cite{hibbert_etal:2002}. The largest difference (about 10\%) was found for the 1404.8~\AA\ transition. We therefore estimate an uncertainty for the ratios of the \siv\ intercombination lines of about 10\%.

Proton excitation among the \oiv~ 2s$^2$ 2p $^2$P and 2s 2p$^2$ $^4$P levels is known to have a small effect on the population of these levels, so we explored whether proton excitation could also affect the \siv\ levels. The CHIANTI model only includes proton rates for the 3s$^2$ 3p $^2$P ground term, however we found estimates of the proton rates to the 3s 3p$^2$ $^4$P levels obtained by \cite{bhatia_etal:1980}. These rates are probably accurate to within a 50\%. We found no significant effects due to the inclusion of these proton rates within the error limits. 

An entire Appendix in \cite{delzanna_badnell:2016} is dedicated to a reassessment of the 
energies of the levels for \siv, and a revision of the rest wavelengths of the lines, for which a large scatter exists in the literature. Of the main four intercombination lines, IRIS only observes two, the $^2$P$_{3/2}$ - $^4$P$_{5/2}$ at 1406.06~\AA, and the $^2$P$_{1/2}$ - $^4$P$_{1/2}$ at 1404.85~\AA.
It is not trivial to measure the rest wavelengths of these lines, because their profiles are often asymmetric and mostly redshifted in the solar atmosphere. \cite{delzanna_badnell:2016} adopted the wavelengths from Skylab observations at the limb, as reported by \cite{sandlin_etal:1986}. Their accuracy is about 0.005~\AA. The first line was directly observed at 1406.059$\pm 0.005$~\AA. The wavelength of the second line was obtained from the 
1423.885$\pm 0.005$~\AA\ decay to the 3s$^2$ 3p $^2$P$_{3/2}$, measured by \cite{sandlin_etal:1986}, and a ground level 3s$^2$ 3p $^2$P separation of 951.4 Kaysers, from measurements of the forbidden line in planetary nebulae, see e.g. \cite{feuchtgruber_etal:1997}. Finally, we note that the NIST database gives wavelengths of 1406.009 and 1404.808~\AA, i.e. for the blended line gives the same wavelength as the \oiv\ line (see below).

\subsection{\oiv}
In this section we discuss the atomic data used for \oiv.

Diagnostic lines from \oiv\ ions have been used for a long time to measure 
electron number densities (see, e.g. \cite{Flower75,feldman_doschek:1979}), and
many atomic calculations have been produced over the years.
The most recent scattering calculations (for all the B-like ions) are those 
of the UK APAP network \citep{liang_etal:2012}, which have been 
corrected for an error and distributed 
within the CHIANTI v.8 database \citep{delzanna_etal:2015_chianti_v8}.
We use these data here. 
\cite{liang_etal:2012} performed $R$-matrix calculations 
for electron-impact excitation amongst 204 close-coupling levels (up to $n=4$).
The excitation rates in previous CHIANTI versions (until 7.1) were from 
the $R$-matrix calculations of 
\cite{blum_pradhan:1992}, as reported by \cite{zhang_etal:1994}.
They included only the lowest 15 ($n=2$) close-coupling levels.
Unpublished calculations for $n=3$ levels were added. 
Version 7.1 included new excitation rates calculated with the 
relativistic DARC $R$-matrix code by \cite{aggarwal_keenan:2008}.
These close-coupling calculations included the energetically lowest
 75 levels ($n=2,3$).
The excitation rates to the 
2s 2p$^{2}$ $^{4}$P levels as calculated by \cite{aggarwal_keenan:2008}
and \cite{liang_etal:2012} are quite similar, within a few percent.

For the A-values, CHIANTI v.8 selected those for the lower levels calculated by \cite{correge_hibbert:2004} with the CIV3 code \citep{hibbert:1975} and for the higher levels those calculated by \cite{rynkun_etal_2012} with the multi-configuration Dirac-Hartree-Fock (MCDHF) GRASP2K code \citep{grasp2k}.
Earlier, \cite{tachiev_froese_fisher:2000} carried out Breit-Pauli multiconfiguration Hartree-Fock calculations for the B-like ions. All these A-values are shown in Table~\ref{tab:ne_o_4}, together with those of \cite{liang_etal:2012}. Clearly, there is excellent agreement to well within 10\% among the calculations, with the exception of \cite{liang_etal:2012}. This is because the latter obtained the best target for the scattering calculations, and did not optimise the calculation for the radiative data. We can therefore reasonably assume an overall uncertainty of about 10\% in the intensity ratios of the intercombination lines, as in the \siv\ case. 

As in the \ion{S}{iv} case, several wavelength measurements can be found in the literature. Earlier version of the CHIANTI database had wavelengths for these lines which were close to those obtained from the level energies recommended by \cite{edlen:1983}. We report these wavelengths in Table~\ref{tab:ne_o_4b}. In turn, these values are minor revisions of an earlier work \citep{edlen:1981}, which was mainly based on the experimental observations of \cite{bromander:1969}, who used a very large (3m focal length) spectrograph and measured the wavelengths to better than 0.01~\AA. Since then, better solar rest wavelengths were obtained by \cite{sandlin_etal:1986} from Skylab observations at the limb, 
with an accuracy of about 0.005~\AA.

The CHIANTI wavelengths were revised in version 7.1 \citep{landi_etal:12_chianti_v7.1}, to include the values obtained by \cite{young_etal:2011} using  HST/STIS  RR Tel spectra. This revision caused some confusion in the literature, especially regarding the wavelength of the line blended with \siv, estimated by \cite{young_etal:2011} at 1404.779~\AA. However, it turns out that these revised wavelengths are not consistent with the \cite{bromander:1969} and the \cite{sandlin_etal:1986} ones, which had better accuracy and were very close. Therefore, a new set of wavelengths, consistent with these two sets of experimental data, have been introduced in CHIANTI v.8  \citep{delzanna_etal:2015_chianti_v8}. The estimated wavelength of the $^2$P$_{3/2}$ - $^4$P$_{3/2}$ line is 1404.806~\AA, i.e. very close to the estimated wavelength for the \siv\ transition, 1404.85~\AA. The high-resolution IRIS spectra support these estimates. Indeed, if the two lines were further apart in wavelength, the IRIS spectra would show a broadened profile, which is not observed. 

\section{Observation of the TR spectra in the AR 12356}
\label{Sect:A2}
\begin{figure*}[!htb]
	\centering
		\includegraphics[width=0.65\textwidth]{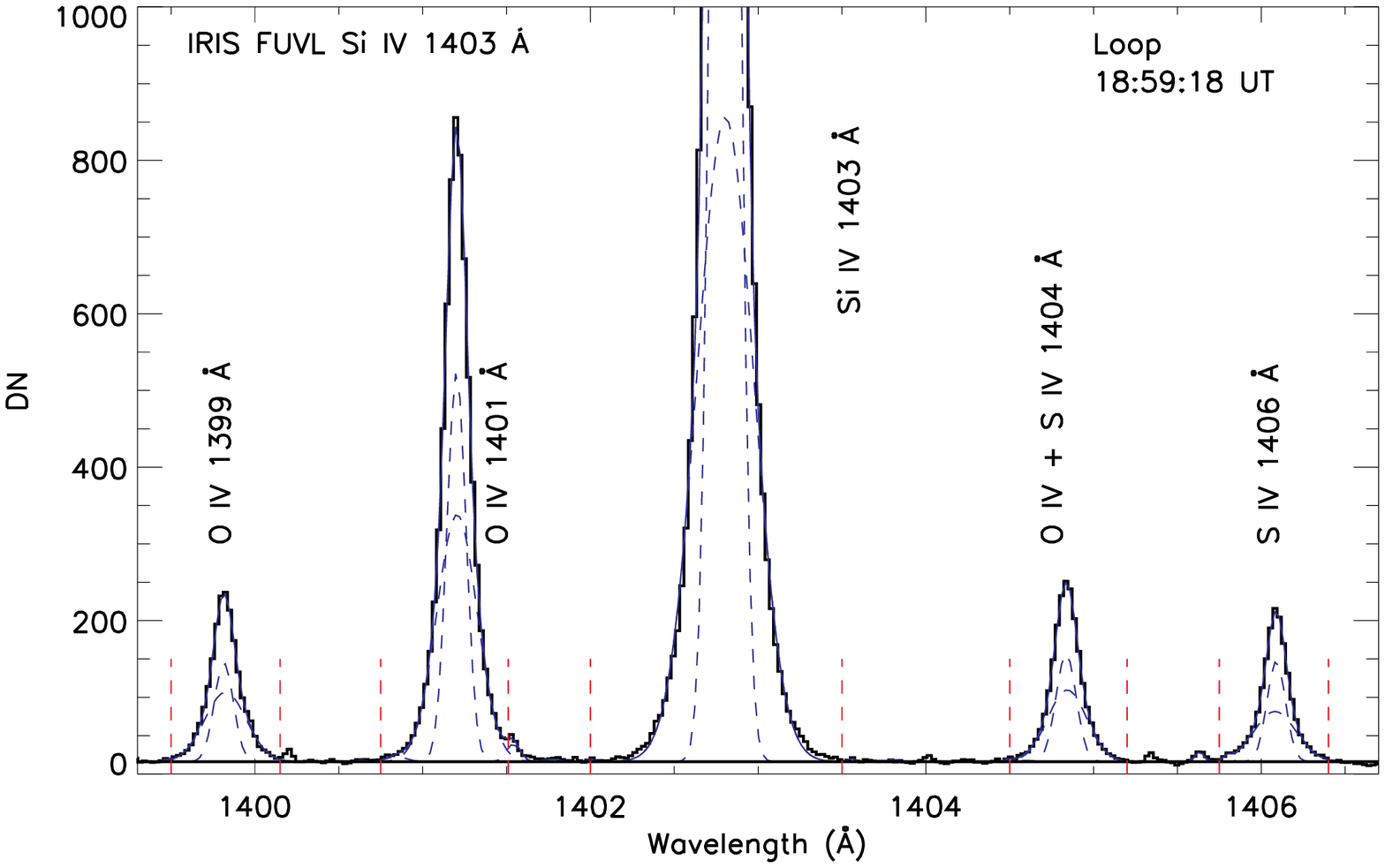} 	
		\includegraphics[width=0.65\textwidth]{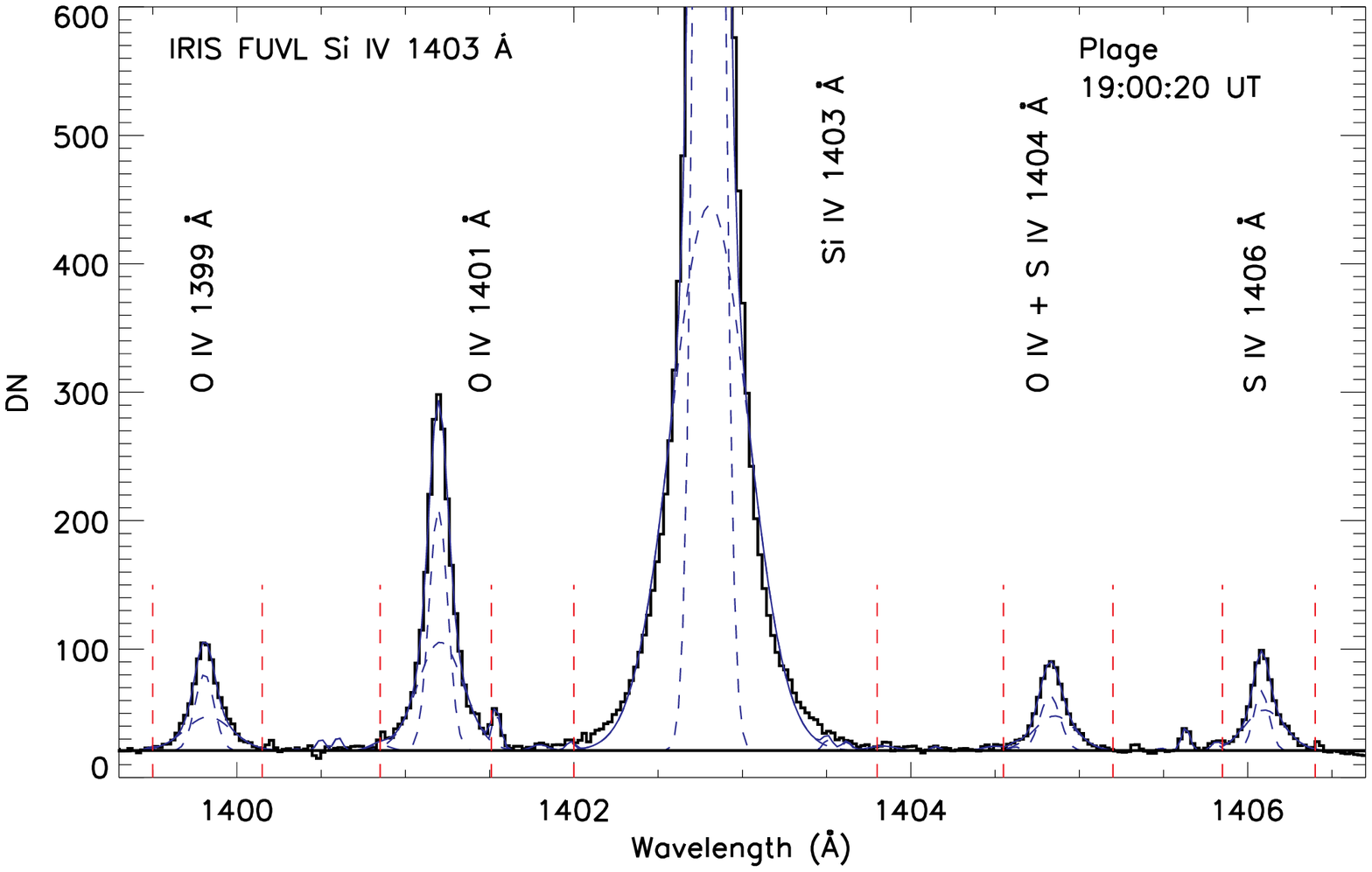} 
		\includegraphics[width=0.65\textwidth]{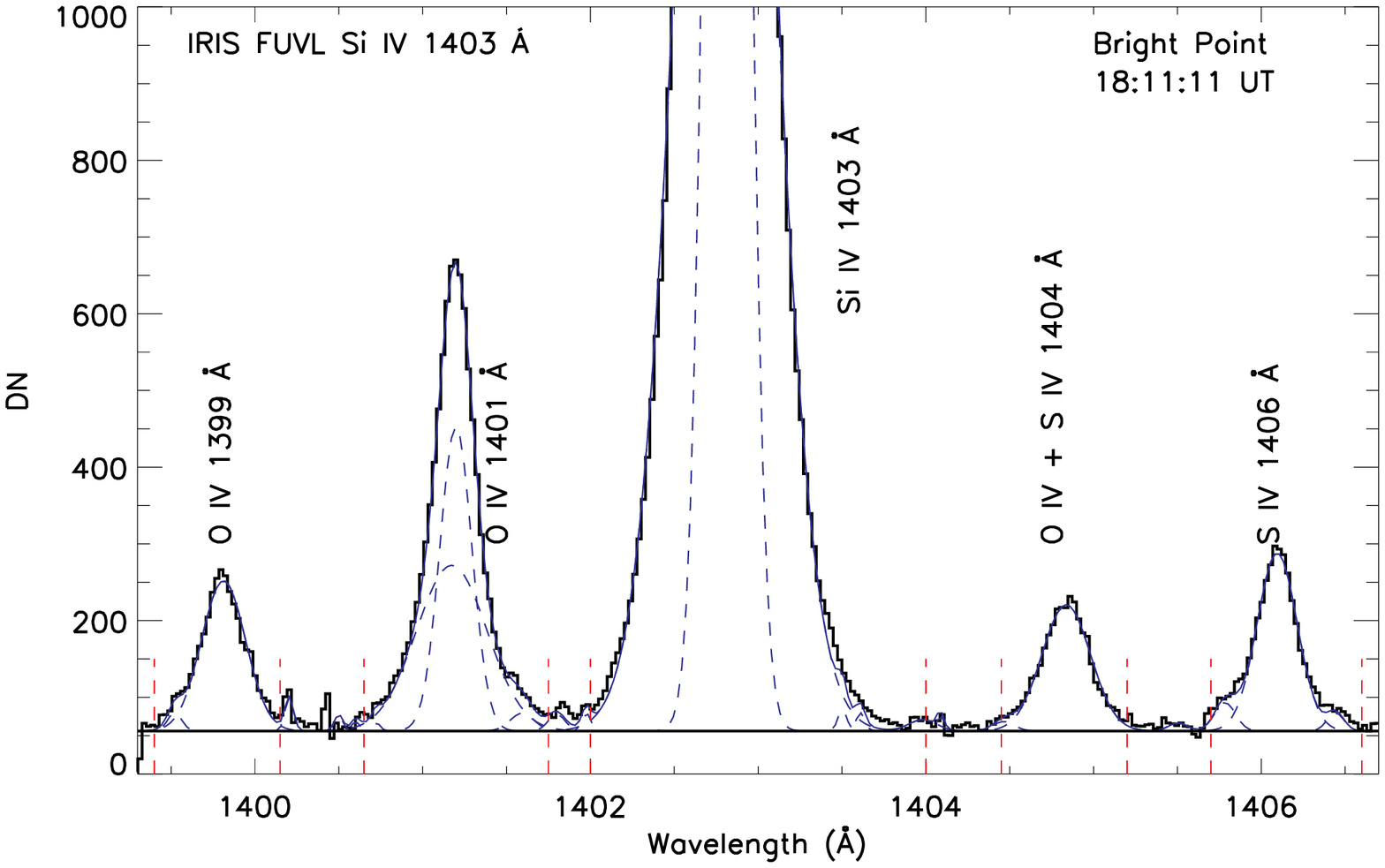} 	
      \caption{Zoomed loop (top), plage (middle) and bright point (bottom) spectra in the AR 12356. The blue dotted lines represent the single Gaussian component from the fit, while the continuous blue line is the sum of these components. We note that the \siiv\ line at 1402.77~\AA~is not properly fitted by two fit components alone. The red vertical dotted lines represent the wavelength interval over which we integrated the total counts for each line. }
      \label{Fig:spectra_appendix}
  \end{figure*}
\begin{figure}
	\centering
	\includegraphics[width=0.45\textwidth]{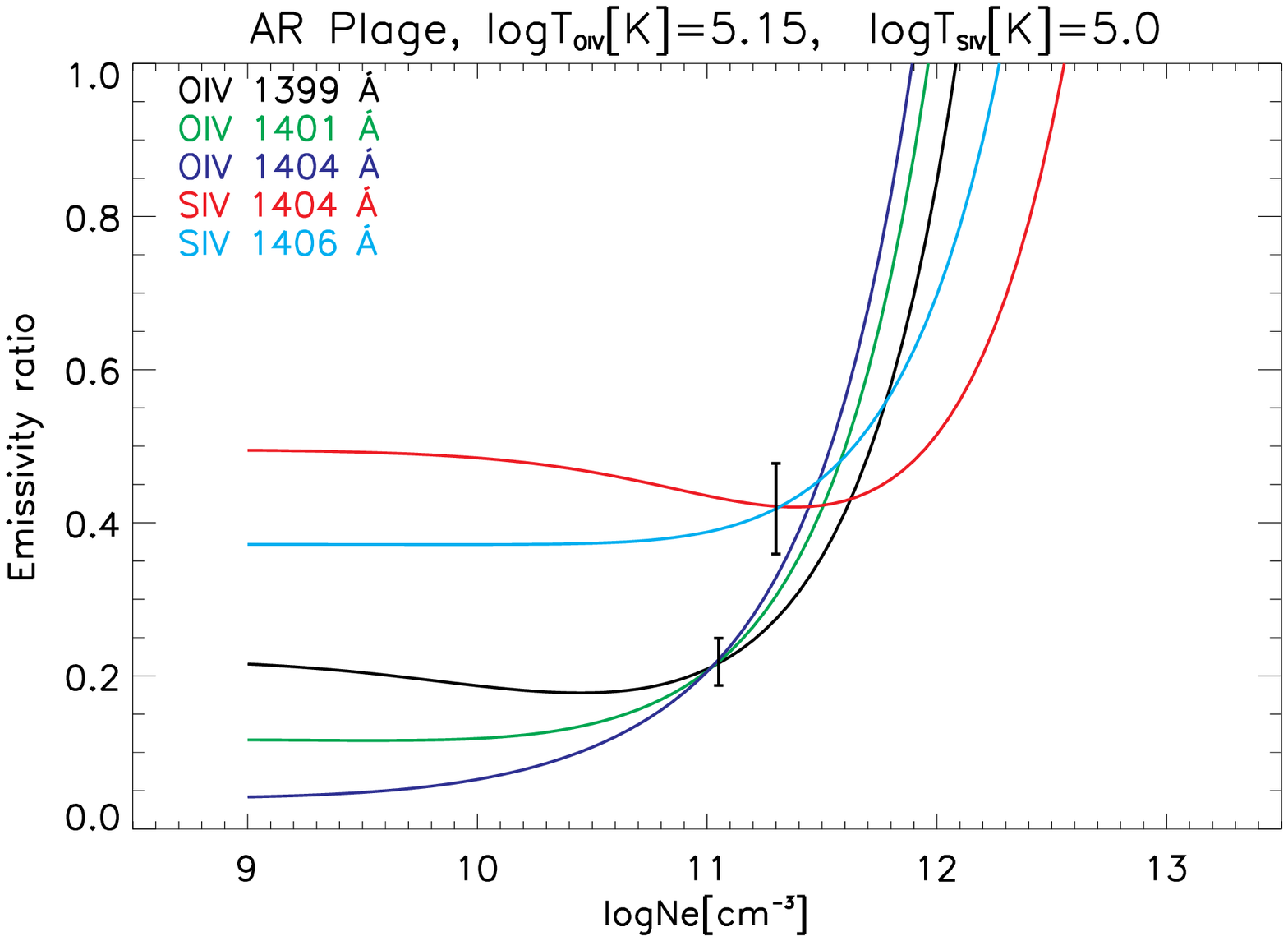} 	
	\includegraphics[width=0.45\textwidth]{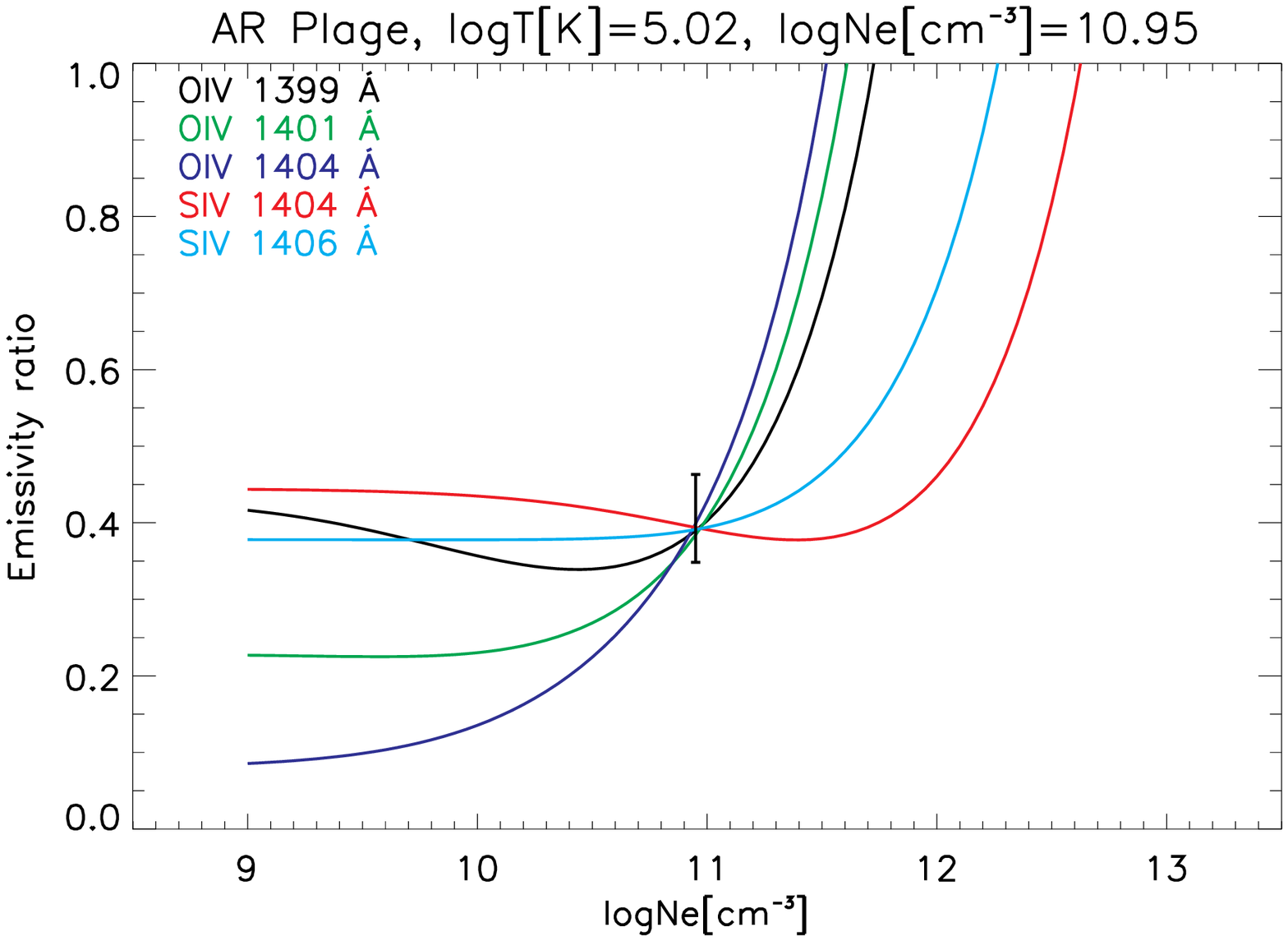} 	
      \caption{Emissivity ratio curves as a function of density for the \oiv~and~\siv~spectral lines observed by IRIS in the "plage" region.Different colours for the curves indicate different spectral lines as described in the legend. For more detail see the caption of the Fig. \ref{Fig:EM_loci_loop}. }
      \label{Fig:EM_loci_plage}
  \end{figure}
\begin{figure}[!ht]
	\centering
	\includegraphics[width=0.45\textwidth]{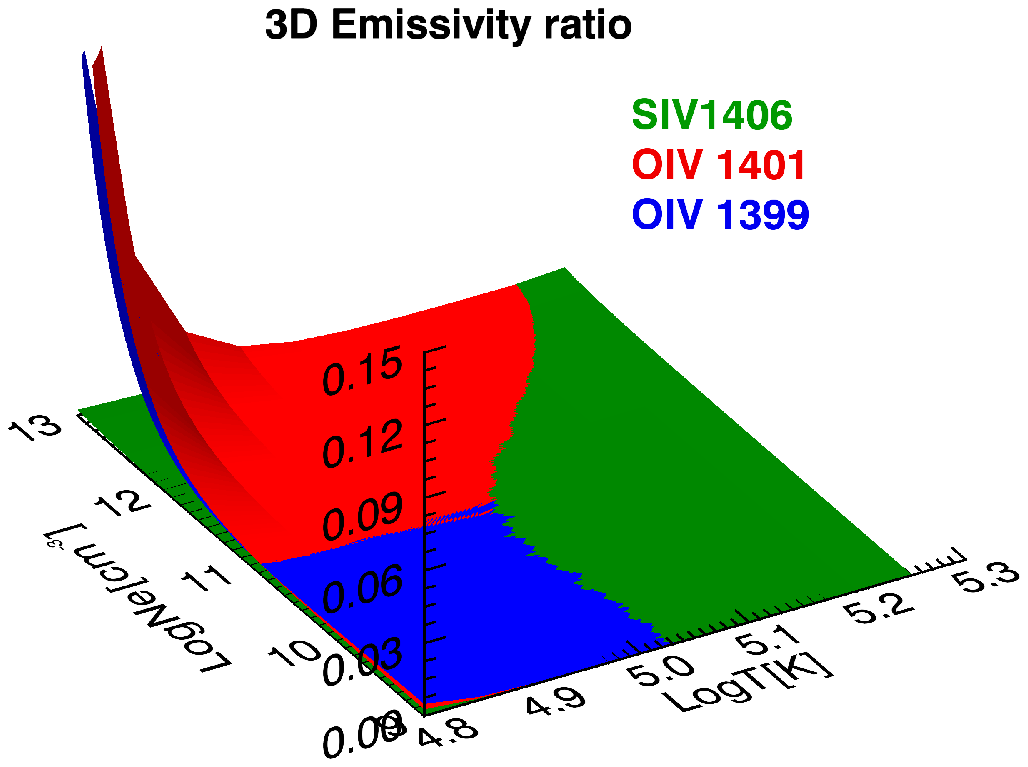} 	
      \caption{3D emissivity ratio as a function of log$N_\textrm{e}$[cm$^{-3}$] and log$T$[K] for the \siv~1406.93~\AA, \oiv~1399.78~\AA~and \oiv~1401.16~\AA~lines observed by IRIS in the "plage" region. Different colours for the surfaces indicate different spectral lines as described in the legend.}
      \label{Fig:surface_plage}
  \end{figure}
\begin{figure}
	\centering
	\includegraphics[width=0.45\textwidth]{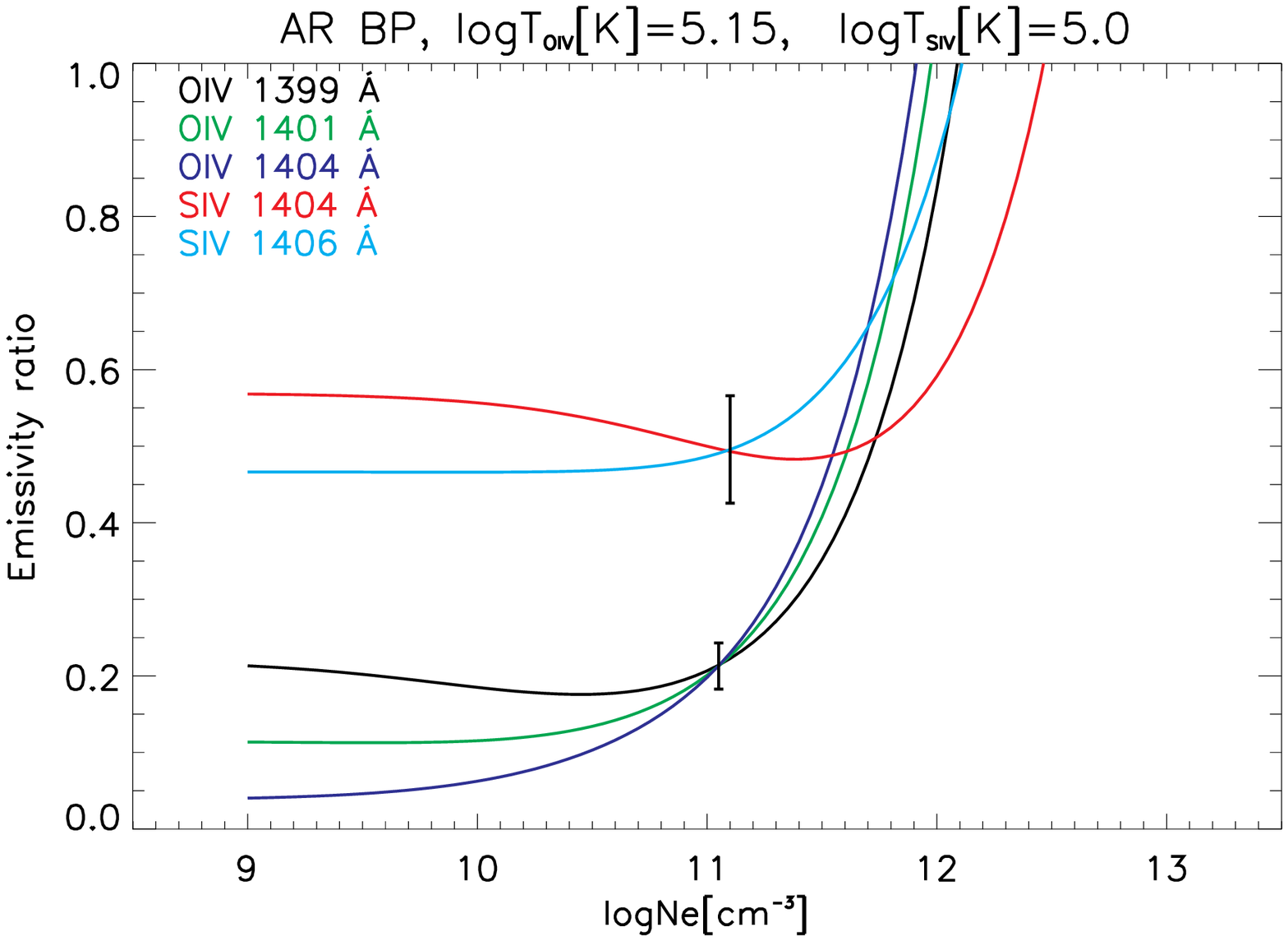} 	
	\includegraphics[width=0.45\textwidth]{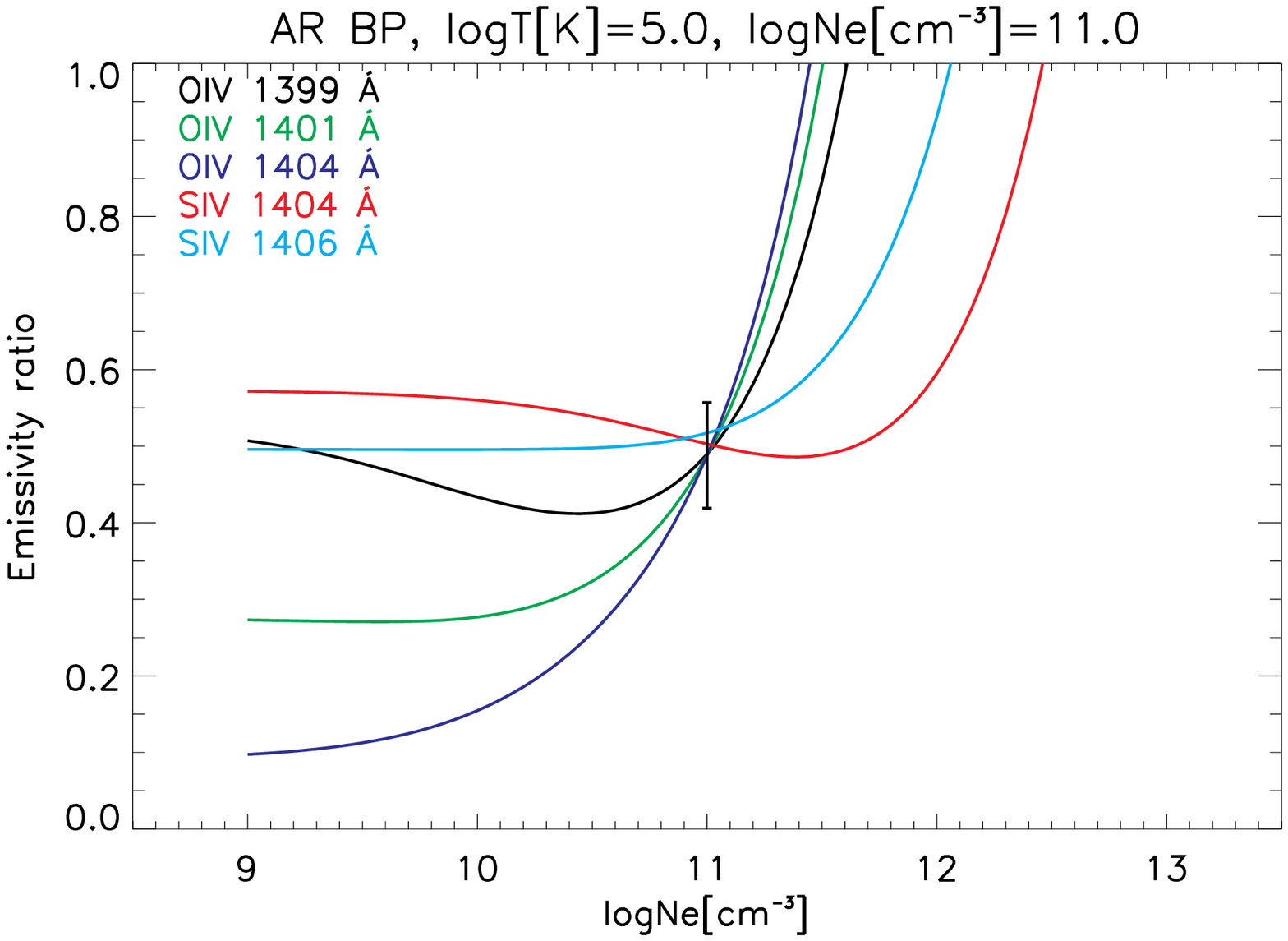} 	
      \caption{Emissivity ratio curves as a function of density for the \oiv~and~\siv~spectral lines observed by IRIS in the "bright point" region. Different colours for the curves indicate different ions as described in the legend}. For more detail see the caption of the Fig. \ref{Fig:EM_loci_loop}. 
      \label{Fig:EM_loci_bright}
  \end{figure}
\begin{figure}[!ht]
	\centering
	\includegraphics[width=0.45\textwidth]{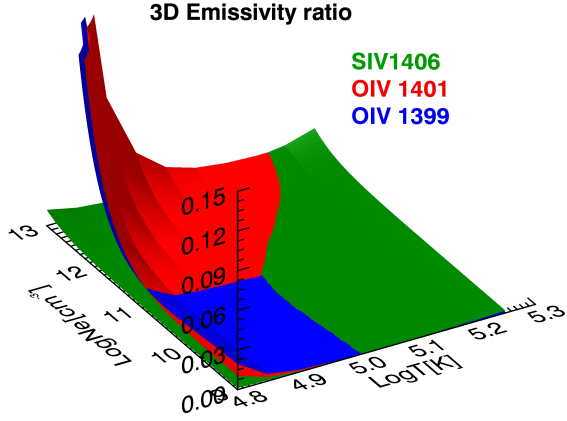} 	
      \caption{3D emissivity ratio as a function of log$N_\textrm{e}$[cm$^{-3}$] and log$T$[K] for the \siv~1406.93~\AA, \oiv~1399.78~\AA~and \oiv~1401.16~\AA~lines observed by IRIS in the "bright point" region. Different colours for the surfaces indicate different ions as described in the legend.}
      \label{Fig:surface_bright}
  \end{figure}

In this section, we provide some additional details for the analysis of the AR plage and bright point spectra shown in Fig. \ref{Fig:TR_maps}, similarly to that presented in Sect. \ref{Sect:4.2} for the loop case. Fig. \ref{Fig:spectra_appendix} shows the individual TR spectra for the AR plage, loop and bright point regions including the double Gaussian fitting. In all cases, the double fit reproduces well the profile of the \oiv\ and \siv\ lines but not of the \siiv\ 1402.77~\AA~line. In Tab. \ref{tab:plage} and \ref{tab:bright} we compare the intensities obtained by using the double fitting (fit) and summing the total counts (sum) for each line under study in the plage and bright point case respectively, as discussed in Sect. \ref{Sect:4.1}. The results of the $ER$ method are shown in Fig. \ref{Fig:EM_loci_plage} and \ref{Fig:surface_plage} for AR plage, and in Fig. \ref{Fig:EM_loci_bright} and \ref{Fig:surface_bright} for the bright point, similar to Sect. \ref{Sect:4.2}. The corresponding density and temperature values estimated are reported in Tab. \ref{tab:densityAR} in the text.  From Fig. \ref{Fig:EM_loci_plage} and \ref{Fig:EM_loci_bright} and Tab. \ref{tab:densityAR} we note that assuming the peak formation temperature for each ion (top panels of Fig. \ref{Fig:EM_loci_plage} and \ref{Fig:EM_loci_bright}) results in electron number densities which are similar to those obtained applying the $ER$ method (bottom panels of Fig. \ref{Fig:EM_loci_plage} and \ref{Fig:EM_loci_bright}). However, in the first case, there is still a large discrepancy between the relative intensity of the \siv\ and \oiv\ lines, as was also observed for the loop and flare cases in Sect. \ref{Sect:4.2} and \ref{Sect:4.3}. This discrepancy seems to be resolved if we assume the same temperature for the \oiv\ and \siv\ plasma, as described in the text.

%

\begin{table}
\begin{center}
 \caption{Line intensities $I_\mathrm{obs}$ in the plage region. }
 \begin{tabular}{llccc}
 \hline\hline\noalign{\smallskip}
 Ion & $\lambda$&$I_\mathrm{obs}$  (fit) & $I_\mathrm{obs}$ (sum) & $\sigma$ \\
&(\AA)& (*)& (*)&\%\\
 \noalign{\smallskip}\hline\noalign{\smallskip}
 \siiv & 1402.77 &-&6230&1.3\%\\
 \oiv &1399.78  &206&202& 7.0\%\\
 \oiv & 1401.16 &637&633& 4.0\%\\
  \siv & 1406.93 &170&166& 7.8\% \\
  \oiv + \siiv & 1404.82&169&169&7.7\%\\
\noalign{\smallskip}\hline
 \end{tabular}
 \tablefoot{(*):$I_\textrm{obs}$ are expressed in phot s arcsec$^{2}$ cm$^{2}$. See caption in Table \ref{tab:loop} for more details.}
 \label{tab:plage}
 \end{center}
 \end{table} 

%
 \begin{table}
\begin{center}
 \caption{Line intensities $I_\mathrm{obs}$ in the bright point region.}
 \begin{tabular}{llccc}
 \hline\hline\noalign{\smallskip}
 Ion & $\lambda$&$I_\mathrm{obs}$  (fit) & $I_\mathrm{obs}$ (sum) & $\sigma$ \\
&(\AA)& (*)&(*)&\%\\
 \noalign{\smallskip}\hline\noalign{\smallskip}
 \siiv & 1402.77 &-&24216&$<$ 1\%\\
 \oiv &1399.78  &734&797&8.6\% \\
 \oiv & 1401.16 &2382&2470& 3.7\%\\
  \siv & 1406.93 &789&885& 10.8\% \\
  \oiv + \siiv & 1404.82&683&684&3.8\%\\
\noalign{\smallskip}\hline
 \end{tabular}
  \tablefoot{(*):$I_\textrm{obs}$ are expressed in phot s arcsec$^{2}$ cm$^{2}$. See caption in Table \ref{tab:loop} for more details.}
 \label{tab:bright}
 \end{center}
 \end{table} 
 

 \section{Correction in non-Maxwellian collision strengths calculations}
\label{Sect:A3}
 
 In this section we provide the corrections to the collision strengths calculations performed by \citet{Dudik14}. Although the error in the collision strengths for the IRIS lines is only a few per cent, the level population can be significantly affected for high densities and low $\kappa \to 2$, affecting the behaviour of the theoretical \oiv\ density-sensitive ratios $R_1$--$R_3$. Revised density-sensitive ratios are shown in Fig. \ref{Fig:Kappa_ratios} in Appendix \ref{Sect:A3}. For $\kappa$\,=\,2, the ratios are shifted to lower densities by about 0.3 dex compared to the Maxwellian ones.
 
 \begin{figure}[!htbp]
	\centering
	\includegraphics[width=0.45\textwidth]{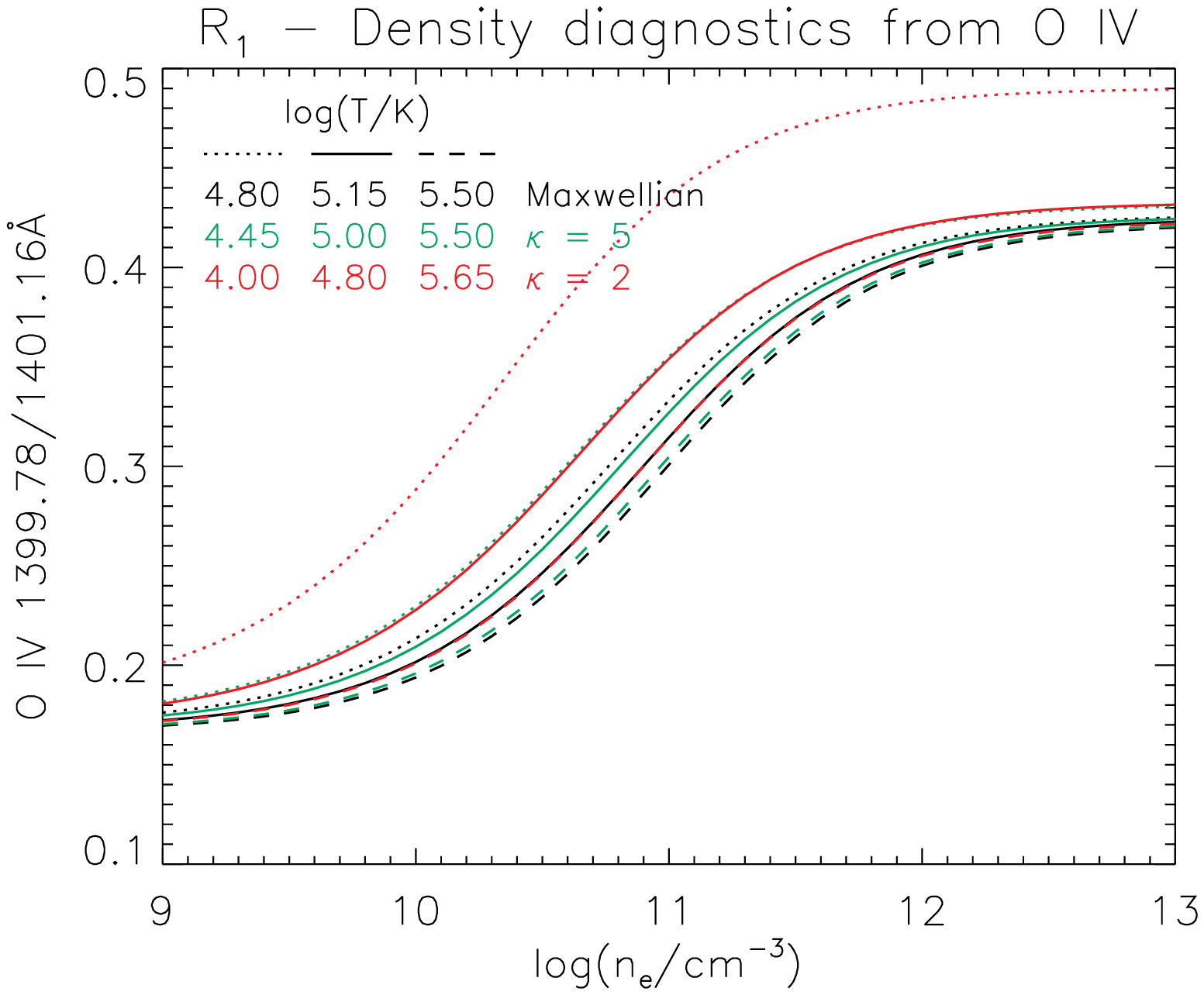} 	
	\includegraphics[width=0.45\textwidth]{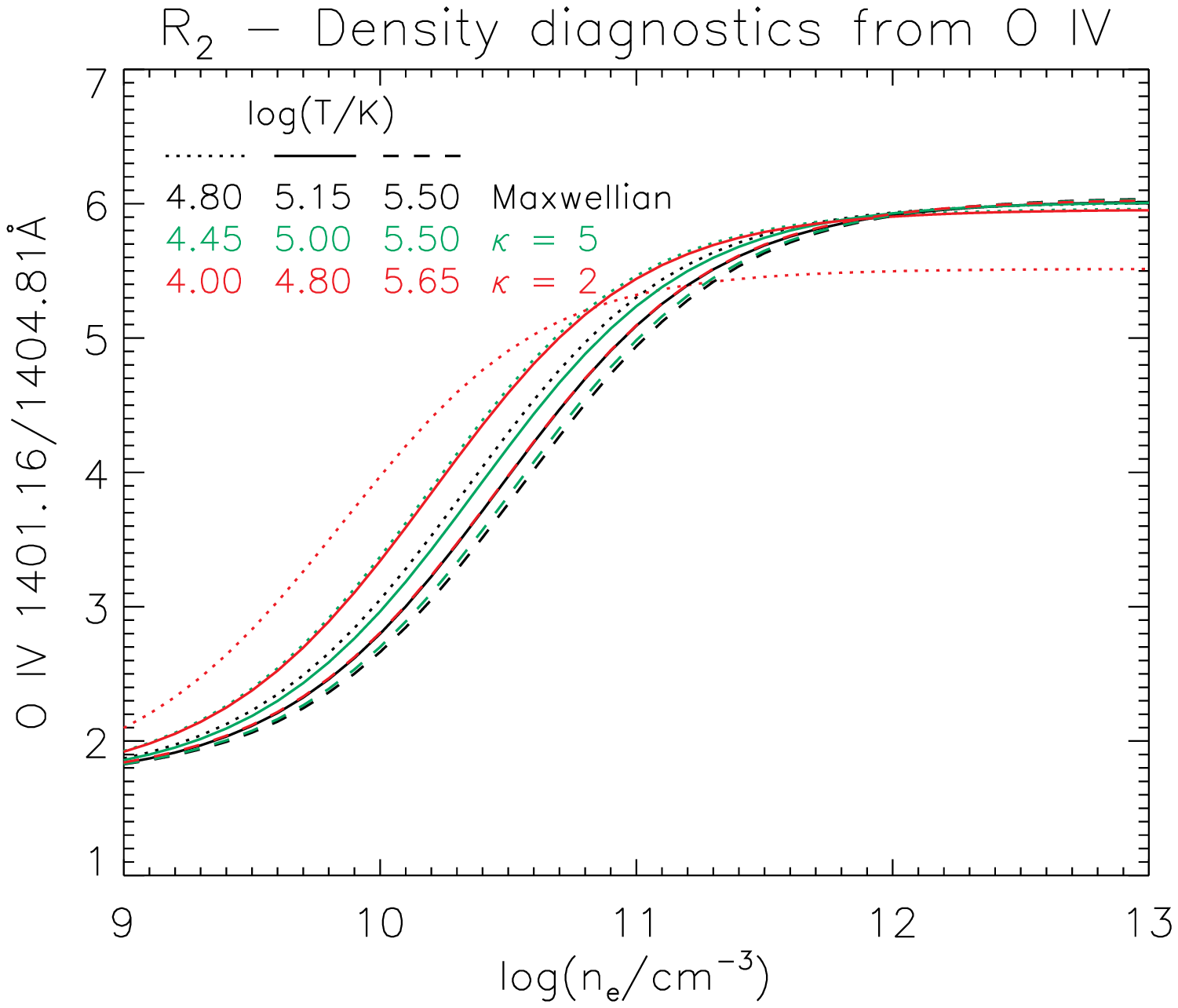} 	
	\includegraphics[width=0.45\textwidth]{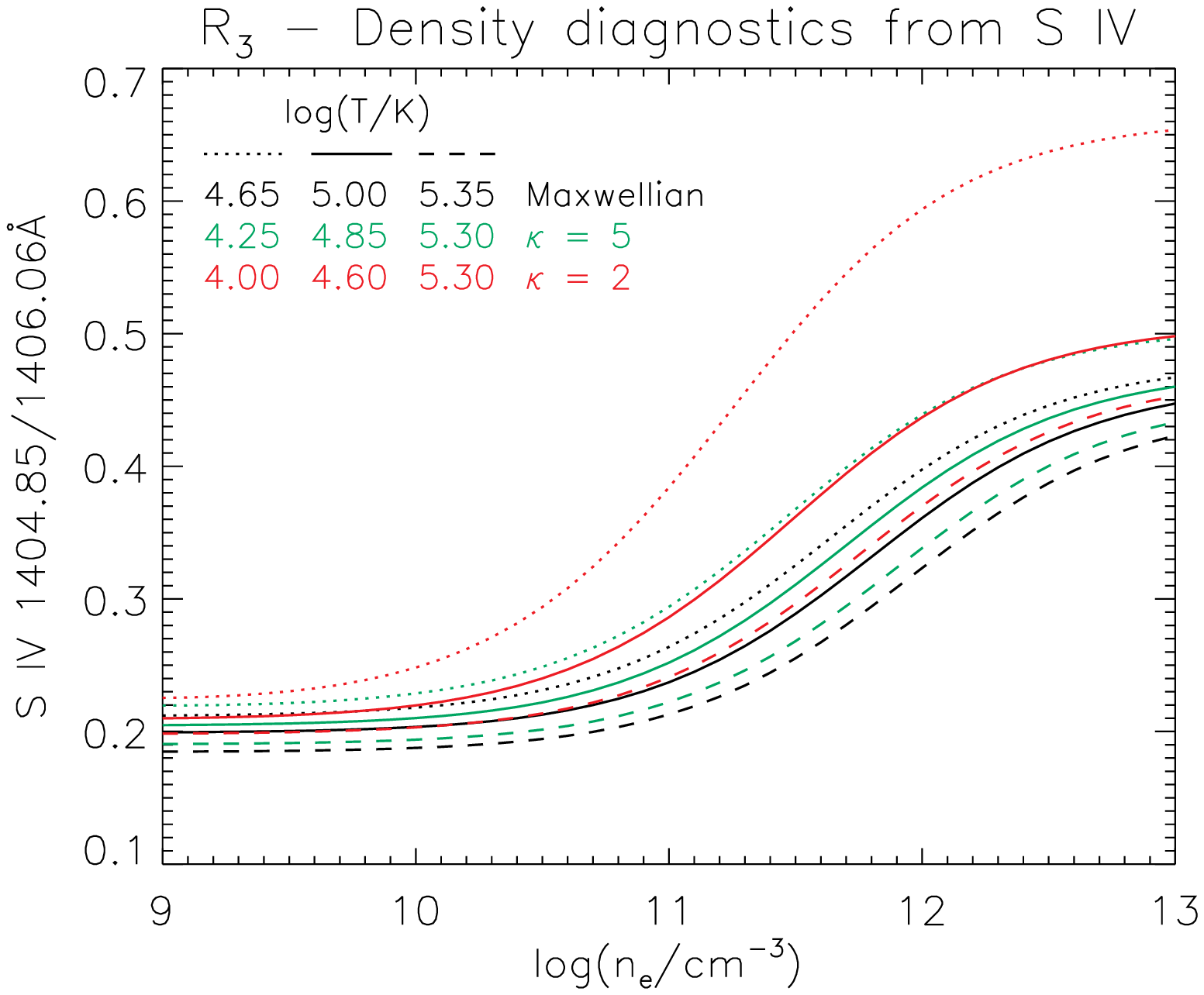} 	
      \caption{Theoretical density diagnostics using the ratios $R_1$--$R_3$ for the non-Maxwellian $\kappa$-distributions. Maxwellian ratios are shown in black, while ratios for $\kappa$\,=\,5 and 2 are shown in green and red, respectively. Three lines for each distribution are shown, with the full lines showing the ratio at the respective peak of the relative ion abundance, while the dotted and dashed ones correspond to temperatures where the relative ion abundance is 0.01 of the maximum value.}
      \label{Fig:Kappa_ratios}
  \end{figure}

\end{appendix}

\bibliographystyle{aa}

\bibliography{arxiv}

\end{document}